\def\year{2021}\relax
\documentclass[letterpaper]{article} 
\usepackage{aaai21}  
\usepackage{times}  
\usepackage{helvet} 
\usepackage{courier}  
\usepackage[hyphens]{url}  
\usepackage{graphicx} 
\usepackage{natbib}  
\usepackage{caption} 
\usepackage{mathrsfs}
\usepackage{epstopdf}
\usepackage{subcaption}
\usepackage{mathtools}
\usepackage{bm}
\usepackage{amsmath,amsfonts,amssymb,amscd,amsthm,xspace}
\usepackage{algorithm}
\usepackage{algorithmic}

\theoremstyle{plain}
\newtheorem{theorem}{Theorem}
\newtheorem{lemma}{Lemma}

\newcommand{\vect}[1]{\boldsymbol{\mathbf{#1}}}
\title{A Hybrid Bandit Framework for Diversified Recommendation}
\author {

        Qinxu Ding,\textsuperscript{\rm 1}
        Yong Liu,\textsuperscript{\rm 1,2}
        Chunyan Miao,\textsuperscript{\rm 3}
        Fei Cheng,\textsuperscript{\rm 4}
        Haihong Tang\textsuperscript{\rm 4}\\
}
\affiliations {
    \textsuperscript{\rm 1} Alibaba-NTU Singapore Joint Research Institute \\
    \textsuperscript{\rm 2} Joint NTU-UBC Research Centre of Excellence in Active Living for the Elderly (LILY) \\
    \textsuperscript{\rm 3}School of Computer Science and Engineering, Nanyang Technological University\\
    \textsuperscript{\rm 4}Alibaba Group\\
    \{qding001@e., stephenliu@, ascymiao@\}ntu.edu.sg, hongyuan.cf@alibaba-inc.com,
    piaoxue@taobao.com
}

\begin{document}
\maketitle

\begin{abstract}


The interactive recommender systems involve users in the recommendation procedure by receiving timely user feedback to update the recommendation policy. Therefore, they are widely used in real application scenarios. Previous interactive recommendation methods primarily focus on learning users' personalized preferences on the relevance properties of an item set. However, the investigation of users' personalized preferences on the diversity properties of an item set is usually ignored. To overcome this problem, we propose the \underline{L}inear \underline{M}odular \underline{D}ispersion \underline{B}andit (LMDB) framework, which is an online learning setting for optimizing a combination of modular functions and dispersion functions. 
Specifically, LMDB employs modular functions to model the relevance properties of each item, and dispersion functions to describe the diversity properties of an item set.
Moreover, we also develop a learning algorithm, called \underline{L}inear \underline{M}odular \underline{D}ispersion \underline{H}ybrid (LMDH) to solve the LMDB problem and derive a gap-free bound on its $n$-step regret. Extensive experiments on real datasets are performed to demonstrate the effectiveness of the proposed LMDB framework in balancing the recommendation accuracy and diversity.

\end{abstract}

\section{Introduction}
The development of interactive recommender systems (IRS) is necessary to capture user's preferences in time. One of the powerful tools is the multi-armed bandit (MAB) which has been extensively studied and applied in different machine learning tasks~\cite{bubeck2012regret}. MAB models an agent that simultaneously attempts to acquire knowledge and optimizes its decisions based on existing knowledge about the environment, during a sequential decision-making process. Assume there are $m$ arms (or actions) presented to a decision-maker on every of $n$ rounds. The objective of the decision-maker is to maximize the sum of rewards over all the $n$ rounds~\cite{agrawal1988asymptotically}. Contextual bandit is a type of MAB framework, where each arm is represented by a feature vector \cite{chu2011contextual,li2010contextual}. In the literature, contextual bandit has been proven to be powerful for solving large scale interactive recommendation problems~\cite{li2010contextual, wen2015efficient} by adopting the user's timely feedback.

In practice, the user's implicit feedback (e.g., click) is commonly used to build the recommender systems, as it can be easily collected~\cite{shi2014collaborative}. Especially, for online applications, this type of feedback becomes an increasingly important source of data to study the user behaviors~\cite{liu2015boosting,liu2018dynamic,yang2018robust}. However, the implicit feedback does not explicitly describe the user preferences. Thus, it is difficult for non-interactive recommender systems to accurately estimate the user's interests. In IRS, the recommendation agents can interact with the users. They can explore the user's interests by sequentially recommending a set of items to her and collecting her implicit feedback in multiple rounds.

Previous contextual bandit based IRS mainly focus on optimizing the accuracy of recommendation results by learning the user preferences on the relevance features of items~\cite{chu2011contextual,li2010contextual}. They usually ignore the diversity of recommendation results, thus are more likely to result in similar item recommendations. A relevant but redundant item set would make the user feel bored with the recommendation result. On the other hand, although there exist some IRS considering diversity in bandit frameworks~\cite{liu2019bandit,qin2014contextual}, they do not learn the user's personalized preferences on the diversity properties of a set of items. In fact, different users may have different preferences with the diversity properties of an item set. For example, a user with specific interest may prefer a relevant item set than a diverse item set, while a user without specific interest may prefer a diverse item set to explore
her interests.

In this paper, we propose a novel bandit learning framework for IRS, which considers the user's preferences on both the item relevance features and the diversity features of the item set. The objective is to achieve a good trade-off between the accuracy and diversity of recommendation results. Specifically, we use modular functions and dispersion functions to model the relevance features and diversity features, respectively. The dispersion function has been shown to be effective in describing the item set diversity when building the non-interactive recommender system~\cite{sha2016framework}. We employ it in our bandit framework to develop the diversified IRS. In summary, the main contributions made in this work are as follows: (1) we propose a novel bandit learning framework, namely \underline{L}inear \underline{M}odular \underline{D}ispersion \underline{B}andit (LMDB), for diversified interactive recommendation; (2) we propose a \underline{L}inear \underline{M}odular \underline{D}ispersion \underline{H}ybrid (LMDH) algorithm to solve the LMDB problem; (3) we also provide theoretical analysis for the algorithm and derive a gap-free upper bound on its scaled $n$-step regret; (4) we perform extensive experiments on simulated datasets and real datasets to verify the effectiveness of the proposed LMDB framework in balancing the recommendation accuracy and diversity.

\section{Related Work}
\subsection{Diversified Recommendation}

The diversity promoting recommendation methods are usually based on greedy heuristics~\cite{wu2019recent}. For example, the maximal marginal relevance (MMR) method~\cite{carbonell1998use} selects a list of items by maximizing a score function defined as a combination of relevance and diversity metrics. In~\cite{qin2013promoting}, an entropy regularizer which quantifies the posterior uncertainty of ratings of items in a set is derived for promoting the diversity of recommendation results. This entropy regularizer is a monotone submodular function, and there exists a greedy algorithm which is guaranteed to produce a near optimal solution for monotone submodular function within a factor $(1-\frac{1}{e})$ of optimal \cite{nemhauser1978analysis}. Determinantal point process (DPP)~\cite{kulesza2012determinantal} has been studied for diversified recommendation problems~\cite{chen2018fast,wilhelm2018practical,wu2019pd}, and it can also be formulated as a monotone submodular maximization problem~\cite{gillenwater2012near}. Besides these submodular functions, dispersion function that needs less computation has also been studied for diversified recommendation~\cite{gollapudi2009axiomatic,sha2016framework}. The above methods are developed for building the non-interactive recommendation systems. For more details of diversity promoting recommendation methods, you can also refer to ~\cite{chapelle2011intent, vargas2014coverage, puthiya2016coverage,parambath2019re}.

\subsection{Interactive Recommendation}

In the literature, bandit frameworks have been widely studied for building interactive recommender systems. In non-contextual bandit methods, the learner cannot access the arm features. Under this setting, the upper confidence bound (UCB) algorithm is proven to be theoretically optimal~\cite{auer2002finite}. However, without using arm features, the performances of UCB algorithm are quite limited in many applications. In contextual bandit methods, the learner can observe the arm features. LinUCB~\cite{li2010contextual} is a representative contextual bandit algorithm proposed for a personalized news recommendation. It assumes the agent recommends a single arm in each round and focuses on optimizing the recommendation accuracy. For large scale combinatorial multi-armed bandit problems, the performances of non-contextual methods can be suboptimal, because the agents do not use arm features ~\cite{chen2013combinatorial}.
To improve the performances, \cite{wen2015efficient} introduces an efficient learning framework that exploits the arm features and considers the modular function maximization problem.
To support the diversified online recommendation, \cite{qin2014contextual} propose a contextual combinatorial bandit framework based on the entropy regularizer~\cite{qin2013promoting}. In addition, \cite{liu2019bandit} develops a full Bayesian online recommendation framework to balance the recommendation accuracy and diversity, by integrating DPP with Thompson sampling~\cite{chapelle2011empirical}. Overall, these methods are developed based on submodular functions. Due to the inefficiency of computing DPP kernel or entropy regularizer, we find dispersion functions are more suitable for our task. In this work, the proposed LMDB framework considers both relevance features and diversity features of arms, and is formulated as an optimization of the combination of modular functions and dispersion functions.

\section{Problem Formulation}
\label{sec3}

In this section, we firstly describe the properties of our utility function. Throughout this paper, we use personalized item recommendation as our motivating example. Under this setting, utility corresponds to the relevance properties and diversity properties of the recommended item set. Suppose that we recommend a set of items only based on their relevance properties, then the user may be bored with the recommendation results as the items are likely to be very similar. Thus, we should consider both the relevance and diversity properties of the item set to attract users. We formulate this task as a combinatorial optimization problem. Suppose that $E=\{1,\dots,L\}$ is a set of $L$ items, called the ground set. $\mathcal{A} \subseteq \{A \subseteq E: |A|\leq K\}$ is a family of subsets of $E$ with up to $K$ items. Given the user's preferences $\vect{\eta} = [\vect{\theta}^T, \vect{\beta}^T]^T \in \mathbb{R}^{d+m}$, where $\vect{\theta} \in \mathbb{R}^{d}$ are the user's preferences on the relevance properties and $\vect{\beta} \in \mathbb{R}^{m}$ are the user's preferences on the diversity properties. Our objective is to find a set $A \in \mathcal{A}$, which maximizes the following utility function
\begin{equation}
 F(A|\vect{\eta}) = \sum_{i=1}^d \theta_i R_i(A) + \sum_{i=1}^m \beta_i V_i(A),
 \label{ut}
\end{equation}
where $R_i(A)$ is a modular function to capture the item's $i$-th relevance property (e.g., the category of the item) and $V_i(A)$ is a dispersion function to capture the item set's $i$-th diversity property (e.g., Cosine distance of the items in the set). In fact, we assume different distance metrics can describe different diversity properties of the item set. Hence, we have different definitions of the dispersion function $V$ in Eq.~\eqref{ut}.

\textbf{Modular Function.} We firstly introduce the properties of the modular function in Eq.~\eqref{ut}. If $R$ is a modular function mapping the set of items $A$ to real values, then we have
\begin{equation}
    R(A)= \sum_{i\in A}R(i).
    \label{eq1a}
\end{equation}
From Eq.~\eqref{eq1a}, we can get the marginal gain of a modular function for adding an item $a$ to the set $A$,
\begin{equation}
    R(A\cup\{a\})-R(A) = R(a),
\end{equation}
which is independent with other items. It is usually consistent with the recommendation scenario, as the user considers the relevance property of an item is independent with those of other items in a set. Therefore, we define the item $a$'s relevance features as follows,
\begin{equation}
 \Delta_{R}(a|A)= \langle R_{1}(a),\dots, R_{d}(a) \rangle.
 \label{eq5}
\end{equation}
Furthermore, a modular function is monotone, if for any subset $A \subseteq B\subseteq E, R(A) \leq R(B)$. Actually if $R(a)$ is non-negative for any item $a \in E$, then $R$ is a monotone modular function.

\textbf{Dispersion Function.} We secondly introduce the properties of the dispersion functions in Eq.~\eqref{ut}. If $V$ is a dispersion function mapping an item set $A$ to real values, then we have,
\begin{equation}
 V(A)= \sum_{\{i,j\}:i,j\in A}  h(i,j),
 \label{eq2a}
\end{equation}
where $h(i,j)$ can be any distance metric (e.g., Cosine distance) for items $i,j\in A$. From Eq.~\eqref{eq2a}, we can get the marginal gain of a dispersion function for adding an item $a$ to the set $A$,
\begin{equation}
    V(A\cup\{a\})- V(A)=\sum_{j\in A}h(a,j),
\end{equation}
which is dependent with the items in $A$. This is also consistent with the recommendation scenario, as the user considers the diversity property of an item $a$ by comparing it with other items in $A$. Therefore, we define the item a's diversity features as,
\begin{equation}
  \Delta_{V}(a|A)= \langle \sum_{j\in A}h_1(a,j),\dots, \sum_{j\in A}h_m(a,j)\rangle.
  \label{eq9a}
\end{equation}

\textbf{Local Linearity.} One attractive property of $F(A|\vect{\eta})$ is that the incremental gains are locally linear. In particular, the incremental gain of adding $a$ to $A$ can be written as $\vect{\eta}^T \Delta(a|A)$, where
 \begin{equation}
     \Delta(a|A) = \left\langle \Delta_{R}(a|A), \Delta_{V}(a|A)\right\rangle.
 \end{equation}
In other words, $\Delta_{R}(a|A)$ corresponds to the relevance gain by item $a$, and $\Delta_{V}(a|A)$ corresponds to the diversity gain by item $a$, conditioned on the items having already been selected into $A$.

\textbf{Optimization.} Another attractive property of our utility function is that if $F(A|\vect{\eta})$ can be formulated as the combination of a monotone modular function and a dispersion function, there is a greedy algorithm which is guaranteed to produce a near-optimal solution \cite{borodin2012max} within a factor $\frac{1}{2}$.  However, their greedy algorithm is ``non-oblivious'' as it is not selecting the next element with respect to the objective function $F(A|\vect{\eta})$. Therefore, we propose Algorithm \ref{alg:algorithm}, which is a greedy algorithm with the objective function $F(A|\vect{\eta})$ for bandit learning setting. The following theorem describes the approximation property of our algorithm.
\begin{theorem}
\label{theorem1}
Let $E$ be the underlying ground set. For any $A\subseteq E$, let $R_{1}(A),\dots,R_{d}(A)$ be any modular functions, and let $h_{1},\dots, h_{m}$ be any distance metrics for dispersion function. Let $\vect{\theta}\in\mathbb{R}^{d}$ and $\vect{\beta}\in\mathbb{R}^{m}$ be the user's preferences. We write $\vect{\eta}=[\vect{\theta}^{T},\vect{\beta}^{T}]^T$. The object function $F(A|\vect{\eta})$ is defined as Eq.~\eqref{ut}. Let $A^{greedy}$ be the solution computed by the Algorithm \ref{alg:algorithm}, and $A^{*}$ be the optimal solution of the constrained optimization problem $\arg\max_{A: |A|\leq K} F(A|\vect{\eta})$. If $\vect{\theta}^T\langle R_{1}(a),\dots,R_{d}(a)\rangle \geq 0$ for any element $a\in E$, and $\vect{\beta}$ is a non-negative vector, then we have
\begin{equation}
    F(A^{greedy}|\vect{\eta}) \geq \gamma F(A^*|\vect{\eta}),
    \label{app_ratio}
\end{equation}
where $\gamma=\frac{1}{4}$. In other words, the approximation ratio of the greedy algorithm is $\gamma$.
\end{theorem}
Note that a similar condition of the user preferences can be found in \citet{yue2011linear}. The condition is reasonable in practice, as the weighted relevance score of each item should be non-negative for a user. In Section \ref{simulated}, we observe that our approximation ratio is close to 1 in practice, which is significantly better than that suggested in Theorem \ref{theorem1}. The proof of Theorem \ref{theorem1} can be found in the Appendix.

\begin{algorithm}[tb]
\caption{Modular Dispersion Greedy Search}
\label{alg:algorithm}
\begin{algorithmic}[1] 
\STATE \textbf{Input}: User preferences $\vect{\eta}$, modular functions $R_{1},\dots,R_{d}$, distance metrics $h_{1},\dots, h_{m}$, candidate set $E$, integer $K$\\
\STATE \textbf{Output}: Item subset $A\subseteq E$ with $|A|=K$
\STATE $A\leftarrow \emptyset$
\WHILE{$|A|<K$}
\STATE $a \leftarrow \arg\max_{a\in E-A} \vect{\eta}^T \Delta(a|A)$
\STATE $A\leftarrow A\cup \{a\}$
\ENDWHILE
\RETURN $A$
\end{algorithmic}
\end{algorithm}

\section{Linear Modular Dispersion Bandit}

In this section, we present our LMDB framework. Let $E=\{1,2,\dots,L\}$ be a ground set of $L$ items and $K\leq L$ be the number of recommended items.  $\vect{\theta}^*\in \mathbb{R}^d$ and $\vect{\beta}^*\in \mathbb{R}^m$ are the user's preferences on relevance features and diversity features respectively, which are unknown to the learning agent. To simplify the notation, we write $\vect{\eta}^{*} = [\vect{\theta}^{*T}, \vect{\beta}^{*T}]^T$. The relevance properties for each item are given by $R_1, \dots, R_d$ and the distance metrics for diversity properties are given by $h_1,\dots,h_m$, which are known to the learning agent.

Our learning agent interacts with the user as follows. At time $t$, a random subset $E_t\subseteq E$ is presented and the agent recommends a list of $K$ items
$A_t =(a_1^t,\dots,a_K^t) \in \Pi_K(E_t)$, where $a_k^t$ is the $k$-th recommended item and $\Pi_K(E_t)$ is the set of all $K$-permutations of the set $E_t$. The reward of item $a_k$ at time $t$, $w_t(a_k^t)$, is a realization of an independent Bernoulli random variable with mean $\vect{\eta}^{*T}\Delta(a_k^t|\{a_1^t, \dots, a_{k-1}^t\})$. The user examines the list from the first item $a_1^t$ to the last item $a_K^t$ and then provides the feedback (e.g., clicks on or ignores each item). Formally, for any list of items $A_t$, the total rewards $f(A_t,w_t)$ can be written as the sum of rewards at each position,
\begin{equation}
    f(A_t,w_t)=\sum_{k=1}^{|A_t|} w_t(a^t_k).
    \label{eq9}
\end{equation}
From the definitions of Eq.~\eqref{eq9}, we notice that
\begin{equation}
    \mathbb{E}[f(A_t,w_t)]=\sum_{k=1}^{|A_t|}\vect{\eta}^{*T}\Delta(a_k^t|\{a_1^t, \dots, a_{k-1}^t\}).
    \label{eq13}
\end{equation}
Eq.~\eqref{eq13} implies that $\mathbb{E}[f(A_t,w_t)]=F(A_t|\vect{\eta}^*)$ for $F$ defined as in Eq.~\eqref{ut}. Thus, Algorithm \ref{alg:algorithm} can greedily select items to achieve reward within a factor $\gamma$ of the optimal with perfect knowledge of $\vect{\eta}^*$. The goal of the learning agent is to maximize the expected cumulative return in $n$-episodes. The regret is formally defined in Section \ref{theoretical}.

\section{Learning Algorithm}
The expected reward of an item is a combination of a modular function and a dispersion function. Thus, we propose a hybrid optimization algorithm to solve the LMDB problem.

The algorithm can access the item relevance feature $\vect{z}_a$ and diversity feature $\vect{x}_a$. It does not know the user's preference $\vect{\theta}^*$ and $\vect{\beta}^*$, and estimates them through repeated interactions with the user. It also has two tunable parameters $\lambda,~\alpha>0$, where $\lambda$ is the regularization parameter, and $\alpha$ controls the trade-off between exploitation and exploration. The details of the proposed LMDH algorithm are summarized in Algorithm \ref{alg:algorithm2}. At the beginning of each step (line 4), LMDH estimates the user's preference as $\hat{\vect{\theta}}_t$ and $\hat{\vect{\beta}}_t$ based on the observations of previous $t-1$ steps. The estimation can be viewed as a least-squares problem and the derivation of line 4 is based on the matrix block-wise inversion. Due to the space limitation, we omit the derivation process. $\vect{H}_t,\vect{B}_t,\vect{M}_t$ and $\vect{u}_t, \vect{y}_t$ summarize the features and click feedback of all observed items in the previous $t-1$ steps, respectively. In the second stage (lines 5-14), LMDH recommends a list of items sequentially
from the subset $E_t \subseteq E$. The list is generated by the greedy algorithm (i.e., Algorithm \ref{alg:algorithm}), where the reward of an item $a$ is overestimated as lines 8-10 by the upper confidence bound (UCB) \cite{auer2002finite}. In the last stage (lines 15-22), LMDH displays the item list $A_t$ to the user and collects her feedback, which is used to update the statistics. The per-step computational complexity of LMDH is $O(LK(d^2+m^2+md))$. Computing matrix inverse is the main computational cost of Algorithm \ref{alg:algorithm2}. In practice, we update $\vect{H}_t^{-1}$ and $\vect{M}_t^{-1}$ instead of $\vect{H}_t$ and $\vect{M}_t$ \cite{golub2012matrix}, which is $O(m^2+d^2)$. LMDH selects $K$ items out of $L$ in $O(LK(d^2+m^2+md))$, because computing UCB for each item requires $O(d^2+m^2+md)$.

\begin{algorithm}[tb]
\caption{Linear Modular Dispersion Hybrid (LMDH)}
\label{alg:algorithm2}
\begin{algorithmic}[1] 
\STATE \textbf{Inputs}: Parameters $\lambda>0$ and $\alpha>0$\\
\STATE $\vect{H}_1\leftarrow \lambda \vect{I}_d, \vect{u}_1\leftarrow \vect{0}_d, \vect{M}_1 \leftarrow \lambda \vect{I}_m, \vect{y}_1 \leftarrow \vect{0}_m, \vect{B}_1 \leftarrow \vect{0}_{d\times m}$
\FOR{$t=1,2,...,n$}
\STATE $\hat{\vect{\theta}}_t\leftarrow \vect{H}_t^{-1}\vect{u}_t, \hat{\vect{\beta}}_t \leftarrow \vect{M}_t^{-1}(\vect{y}_t-\vect{B}_t^T\hat{\vect{\theta}}_t)$
\STATE $A_t\leftarrow \emptyset$
\FOR{$k=1,2,...,K$}
\FOR{all $a\in E_t\backslash A_t$}
\STATE $\vect{z}_a \leftarrow \Delta_{R}(a|A_t)$, $\vect{x}_a\leftarrow \Delta_{V}(a|A_t)$
\STATE ${v}_a\leftarrow \vect{z}_a^T\vect{H}_t^{-1}\vect{z}_a-2\vect{z}_a^T\vect{H}_t^{-1}\vect{B}_t\vect{M}_t^{-1}\vect{x}_a+\vect{x}_a^T\vect{M}_t^{-1}\vect{x}_a+\vect{x}_a^T\vect{M}_t^{-1}\vect{B}_t^{T}\vect{H}_t^{-1}\vect{B}_t\vect{M}_t^{-1}\vect{x}_a$
\STATE ${\mu}_a \leftarrow \hat{\vect{\theta}}_t^T\vect{z}_a+\hat{\vect{\beta}}_t^T\vect{x}_a+\alpha\sqrt{{v}_a}$
\ENDFOR
\STATE $a_k^t\leftarrow \arg\max_{a\in E_t\backslash A_t}\mu_a$
\STATE $A_t\leftarrow A_t\cup \{a_k^t\}$
\ENDFOR
\STATE Recommend $A_t$ with the order $(a_1^t,\dots,a_K^t)$, and observe the reward $\{w_a\}_{a\in A_t}$
\STATE $\vect{H}_t \leftarrow \vect{H}_t+\vect{B}_t\vect{M}_t^{-1}\vect{B}_t^T$
\STATE $\vect{u}_t\leftarrow \vect{u}_t+\vect{B}_t\vect{M}_t^{-1}\vect{y}_t$
\STATE $\vect{M}_{t+1}\leftarrow \vect{M}_t+\sum_{a\in A_t}\vect{x}_a \vect{x}_a^T$
\STATE $\vect{B}_{t+1}\leftarrow \vect{B}_t+\sum_{a\in A_t}\vect{z}_a \vect{x}_a^T$
\STATE $\vect{y}_{t+1}\leftarrow \vect{y}_t+\sum_{a\in A_t}\vect{w}_a \vect{x}_a^T$
\STATE $\vect{H}_{t+1}\leftarrow \vect{H}_{t}+\vect{z}_a \vect{z}_a^T-\vect{B}_{t+1}\vect{M}_{t+1}^{-1}\vect{B}_{t+1}^T$
\STATE $\vect{u}_{t+1}\leftarrow \vect{u}_{t}-\vect{B}_{t+1}\vect{M}_{t+1}^{-1}\vect{y}_{t+1}$
\ENDFOR
\end{algorithmic}
\end{algorithm}

\section{Regret Analysis}
\label{theoretical}

This section provides the regret analysis for our proposed algorithm LMDH. Let $\gamma=\frac{1}{4}$,  $A^*_t$ be the optimal solution for Eq.~\eqref{eq13} at time $t$, and $A_t$ be the solution selected from Algorithm \ref{alg:algorithm2}. The $\gamma$-scaled $n$-step regret is defined as follows,
\begin{equation}
    R^{\gamma}(n)=\sum_{t=1}^n\mathbb{E}[F(A^*_t|\vect{\eta}^*)-F(A_t|\vect{\eta}^*)/\gamma],
    \label{eq14}
\end{equation}
where $1/\gamma$ accounts for the fact that $A_t$ is a $\gamma$-approximation at any time $t$. Similar definitions of this scaled regret can be found in \cite{chen2016combinatorial, wen2017online}. This definition is reasonable, because it is computationally inefficient to get the optimal solution $A^*_t$ of this problem, even for the offline variant setting. We now state our main result, which essentially bounds the greedy regret.

\begin{theorem}
\label{theorem2}
Under LMDH, for any $\lambda \geq 0$, any $\delta\in(0,1)$, any $\|\vect{\eta}^*\|_2\leq 1$, and any
\begin{equation}
\begin{aligned}
 \alpha \geq  &\sqrt{(d+m)\log \left[1+\frac{nK}{(d+m)\lambda}\right]+2\log\left(\frac{1}{\delta}\right)}\|\vect{\eta}^*\|_2\\
 &+\lambda^{\frac{1}{2}}
\end{aligned}
\label{eq13a}
\end{equation}
in Algorithm \ref{alg:algorithm2}, we have
\begin{equation}
     R^\gamma(n) \leq \frac{2\alpha K}{\gamma}\sqrt{\frac{n(d+m)\log \left[1+\frac{nK}{(d+m)\lambda}\right]}{\lambda\log(1+\frac{1}{\lambda})}}+nK\delta.
\end{equation}
\end{theorem}

Note that if we choose $\lambda=1,~\delta=\frac{1}{nK},$ and $\alpha$ as the lower bound specified in Eq.~\eqref{eq13a}, then the regret bound in Theorem \ref{theorem2} is $\tilde{O}(K(d+m)\sqrt{n}/\gamma)$. The factor $\sqrt{n}$ is considered near-optimal in gap-free regret bounds. The factor $d+m$ is the number of features, which is standard in linear bandits \cite{abbasi2011improved}. The factor $\frac{1}{\gamma}$ is due to the fact that $A_t$ is a $\gamma$-approximation. The factor $K$ is because of the number of recommended items for each trial. The proof of Theorem \ref{theorem2} can be found in the Appendix.

\section{Experiments}
\label{experiments}
In this section, we perform two kinds of numerical experiments to verify the performance of our learning algorithm. In the first experiment, we show our algorithm outperforms other baseline methods for the task of balancing the accuracy and diversity of the recommendation result on real-world datasets. In the second experiment, we compare the regret of our algorithm with other baseline methods to verify our Theorem \ref{theorem2} on simulated datasets and validate the approximation ratio of Algorithm \ref{alg:algorithm}.

\subsection{Experiments on Real-world Datasets}
\label{real-data}

The experiments are performed on the following datasets: Movielens-100K, Movielens-1M\footnote{https://grouplens.org/datasets/movielens/}, and Anime\footnote{https://www.kaggle.com/CooperUnion/anime-recommendations-database}. Movielens-100K contains 100,000 ratings given by 943 users to 1682 movies, and Movielens-1M contains 1,000,209 ratings given by 6,040 users to 3,706 movies. 
Anime contains 7,813,737 ratings given by 73,515 users to 11,200 animes. 
Following~\cite{liu2019bandit}, we keep the ratings larger than 3 as positive feedback on Movielens-100K and Movielens-1M, and keep the ratings larger than 6 as postive feedback on Anime. We use the unbiased offline evaluation strategy~\cite{li2011unbiased} to evaluate different recommendation methods. Firstly, we randomly split each dataset into two non-overlapping sets. Specifically, $80\%$ of the users are used for training, and the remaining $20\%$ users are used for testing. After that, BPRMF~\cite{rendle2012bpr} is employed to learn the embeddings of all items from the training data. Empirically, we set the dimensionality of the item embeddings to 10. The item embeddings are then normalized into a bounded space $[-1,1]^{10}$, which is considered as the relevance features $\vect{z}_a$ of an item $a$. For fair comparison, 
we only use the average Cosine distance as the distance metric of the dispersion function, which is defined by $\frac{2}{|A_t|(|A_t|-1)}(1-sim(\vect{z}_i, \vect{z}_j))$, where $sim(\vect{z}_i, \vect{z}_j)$ denotes the Cosine similarity between the relevance features of two items $i$ and $j$ in $A_t$, and $|A_t|$ is the size of the item set $A_t$. For each trial $t$, the candidate item set $E_t$ is generated by removing the recommended items of the last $t-1$ steps from the total item set. 
The statistics of the experimental datasets are summarized in Table 1 in Appendix.

\begin{figure*}[t!]
     \centering
     \begin{subfigure}[b]{0.33\textwidth}
         \centering
         \includegraphics[width=\textwidth]{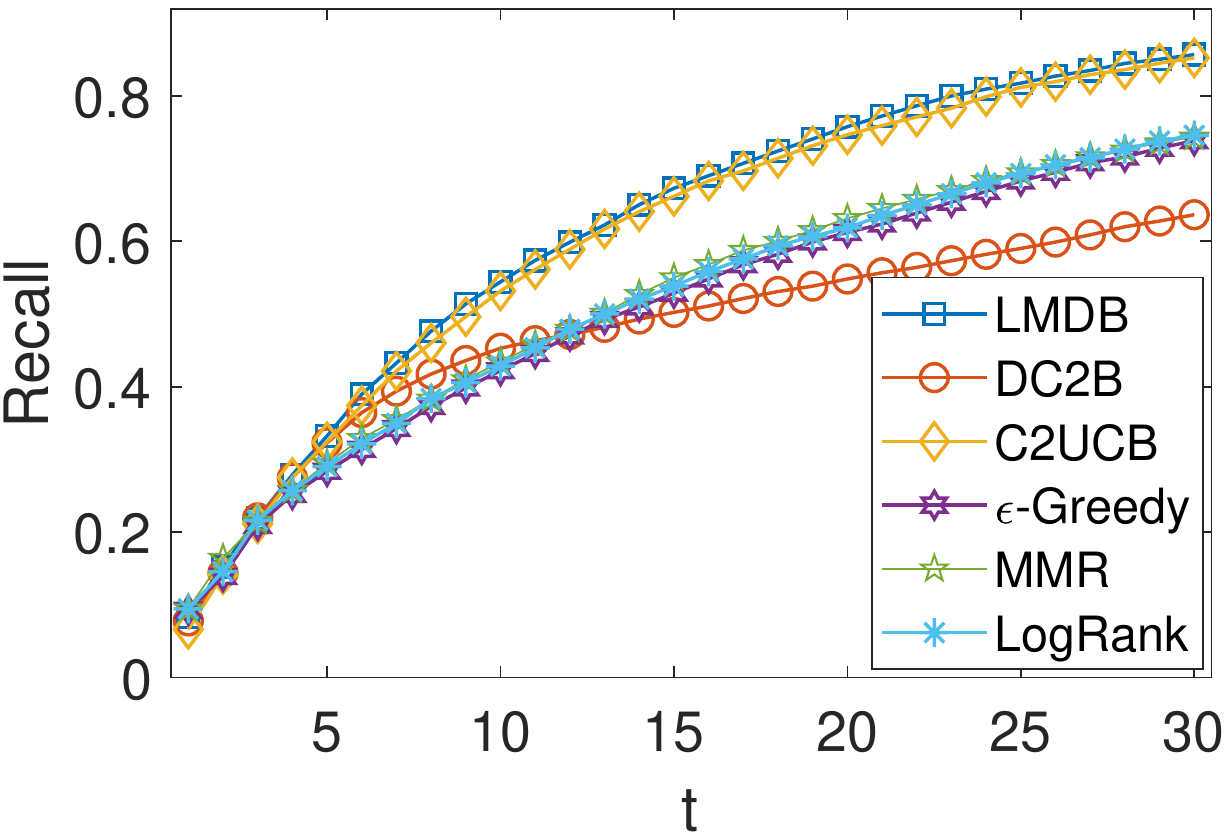}
         \caption{Movielens-100K}
         \label{fig1a}
         \end{subfigure}\hfill\begin{subfigure}[b]{0.33\textwidth}
         \centering
         \includegraphics[width=\textwidth]{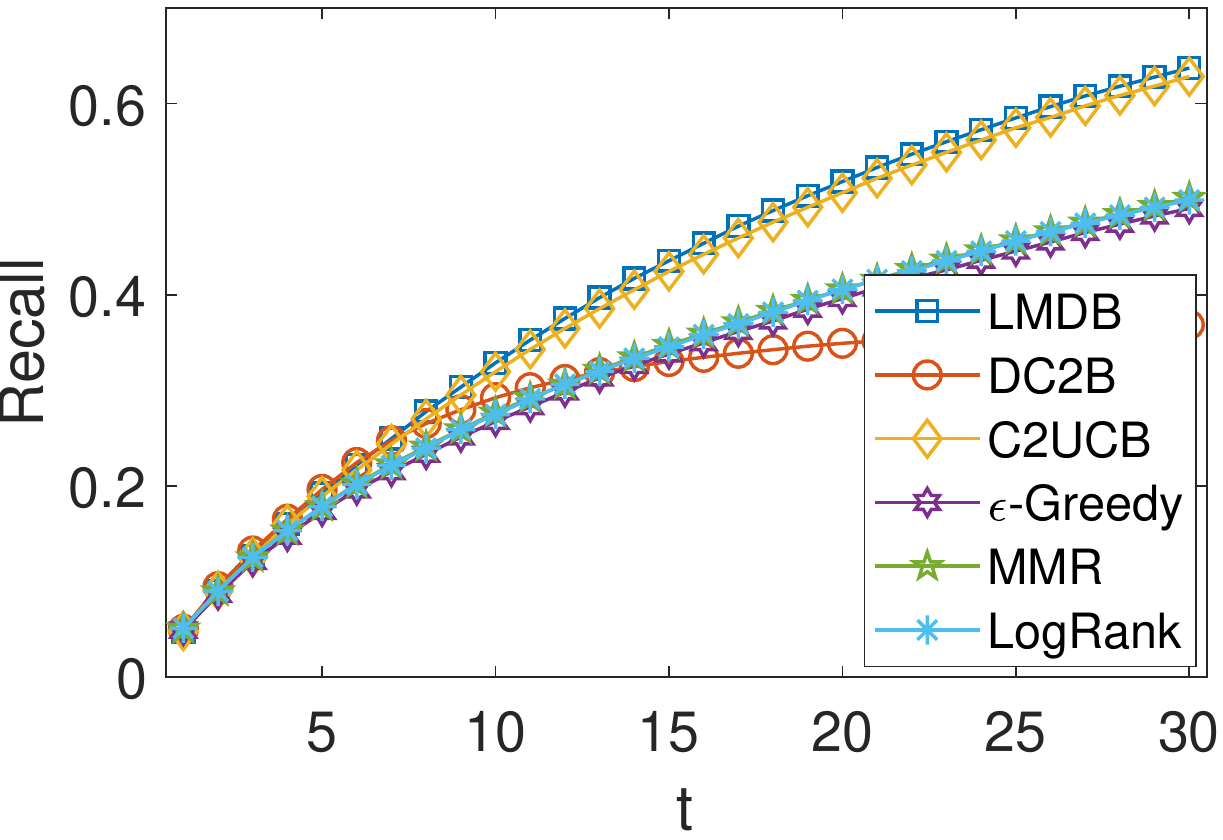}
         \caption{Movielens-1M}
         \label{fig1b}
     \end{subfigure}\hfill \begin{subfigure}[b]{0.33\textwidth}
         \centering
         \includegraphics[width=\textwidth]{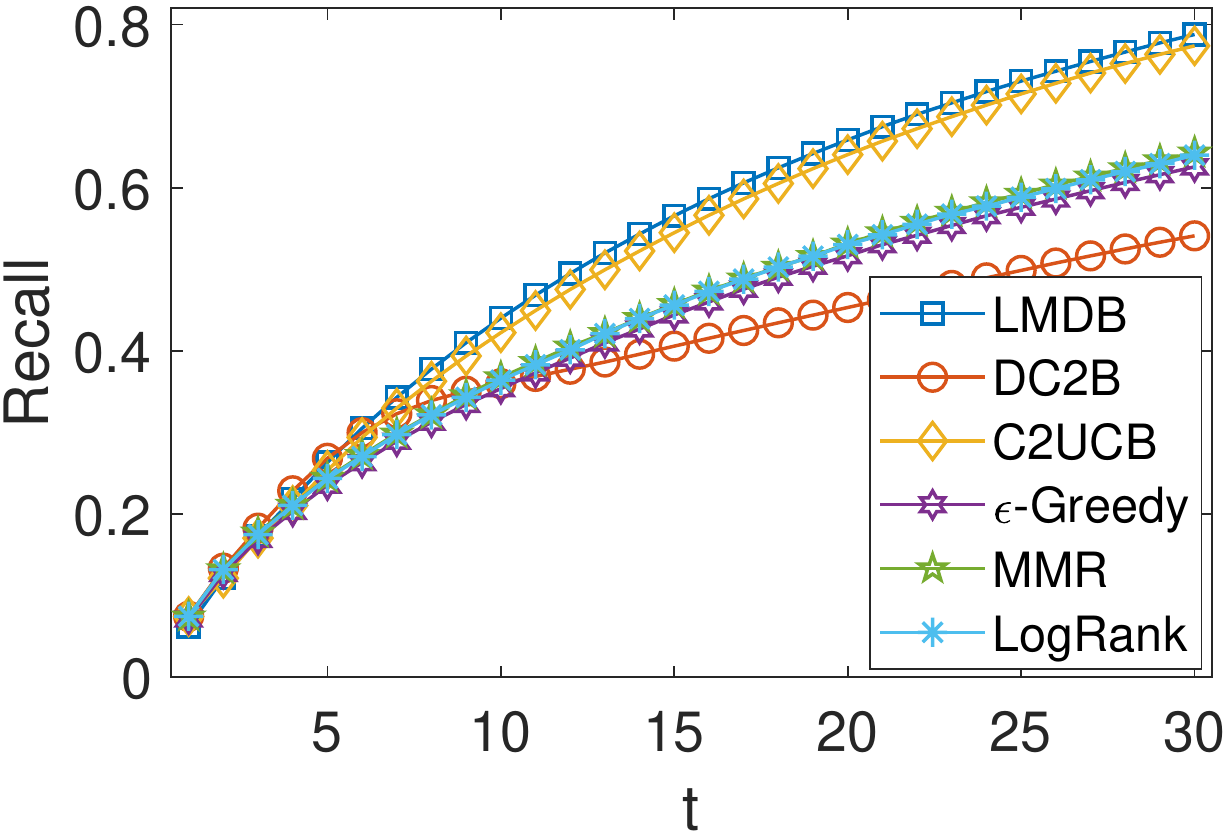}
         \caption{Anime}
         \label{fig1c}
     \end{subfigure}
        \caption{The recommendation performances of different methods measured by Recall.}
        \label{fig1}
\end{figure*}
\begin{figure*}[t!]
     \centering
     \begin{subfigure}[b]{0.33\textwidth}
         \centering
         \includegraphics[width=\textwidth]{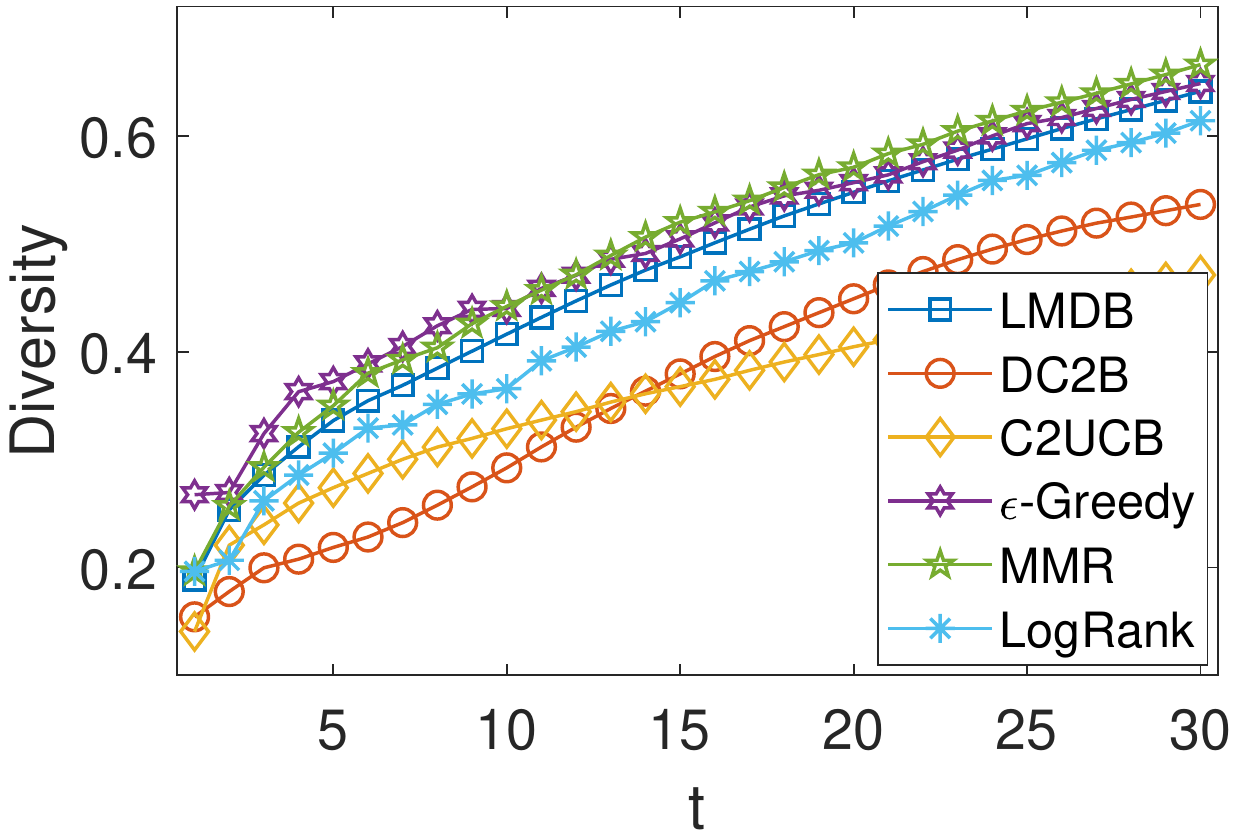}
         \caption{Movielens-100K}
         \label{fig2a}
     \end{subfigure}\hfill\begin{subfigure}[b]{0.33\textwidth}
         \centering
         \includegraphics[width=\textwidth]{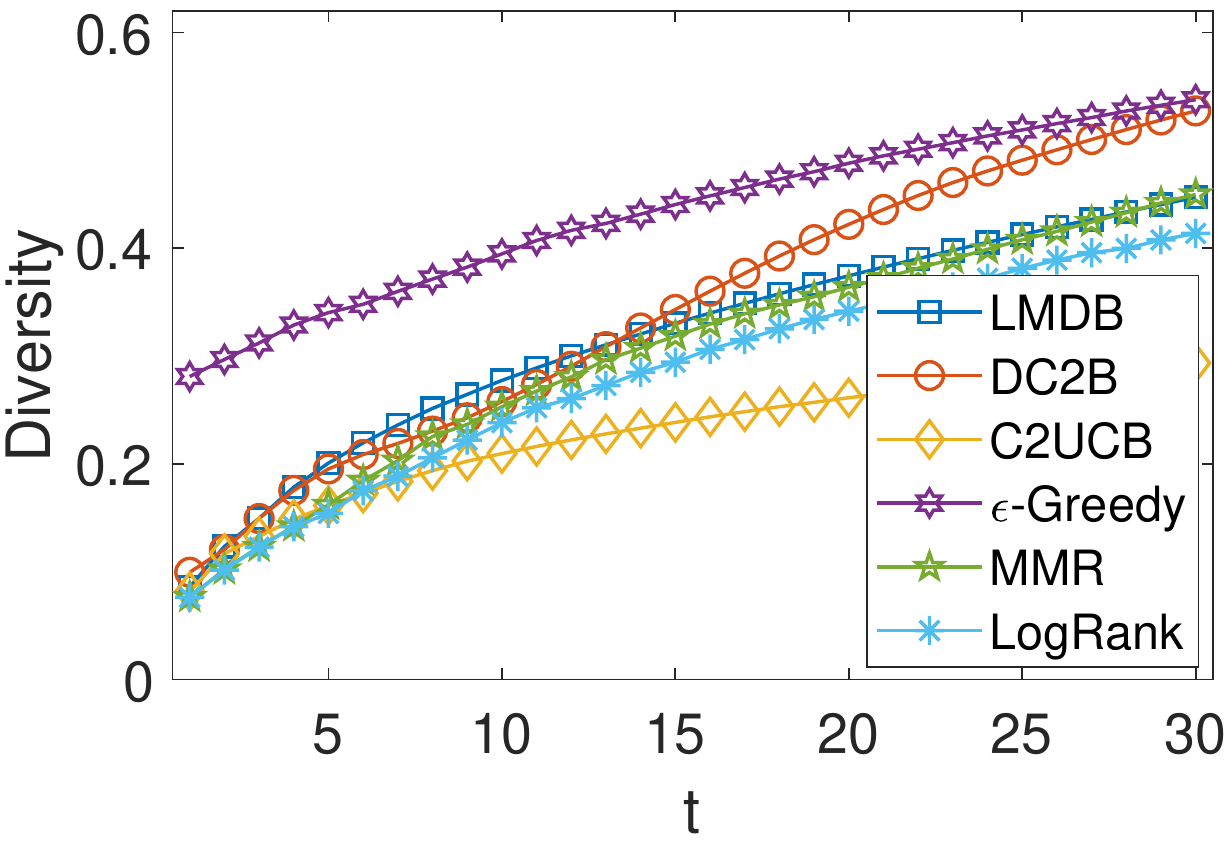}
         \caption{Movielens-1M}
         \label{fig2b}
     \end{subfigure}\hfill\begin{subfigure}[b]{0.33\textwidth}
         \centering
         \includegraphics[width=\textwidth]{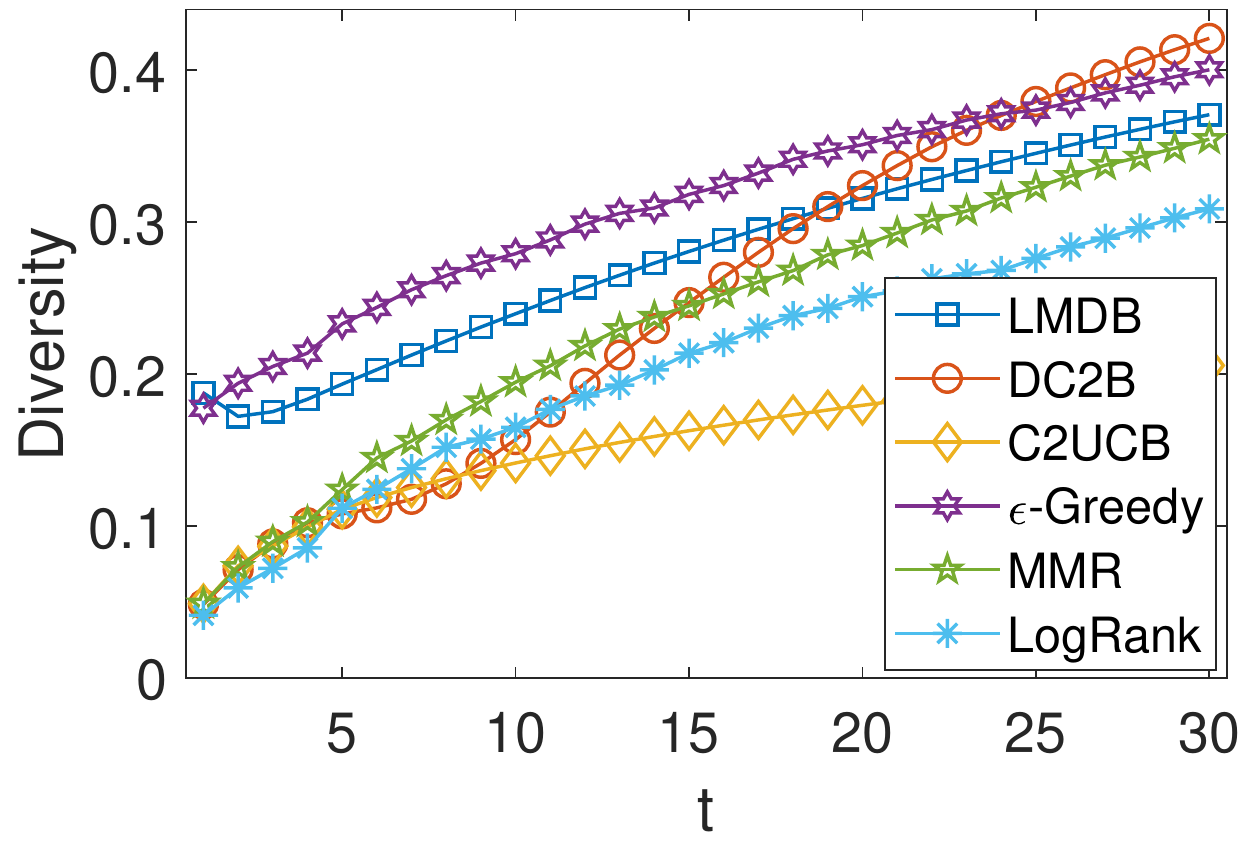}
         \caption{Anime}
         \label{fig2c}
     \end{subfigure}
     \caption{The recommendation performances of different methods measured by Diversity.}
        \label{fig2}
\end{figure*}
\begin{figure*}[t!]
     \centering
     \begin{subfigure}[b]{0.33\textwidth}
         \centering
         \includegraphics[width=\textwidth]{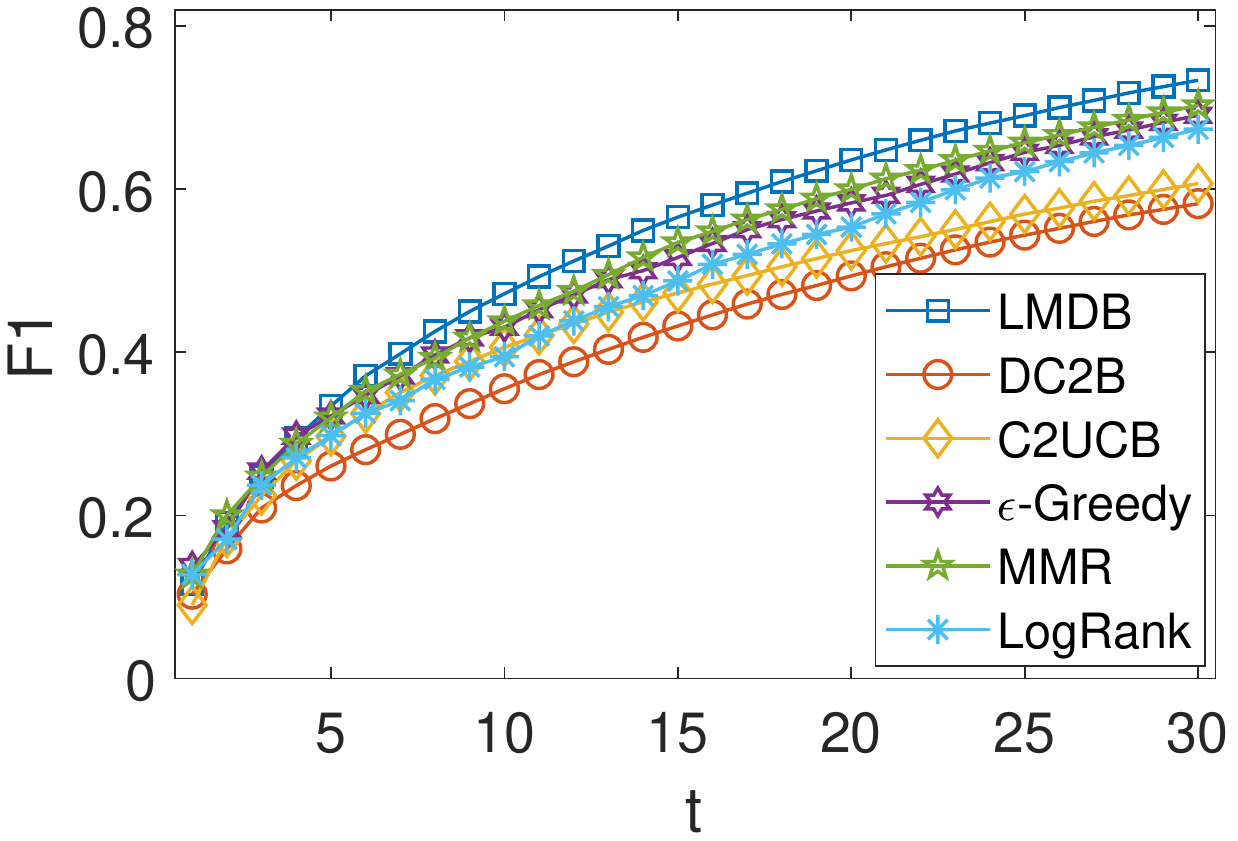}
         \caption{Movielens-100K}
         \label{fig:f1score1}
     \end{subfigure}\hfill\begin{subfigure}[b]{0.33\textwidth}
         \centering
         \includegraphics[width=\textwidth]{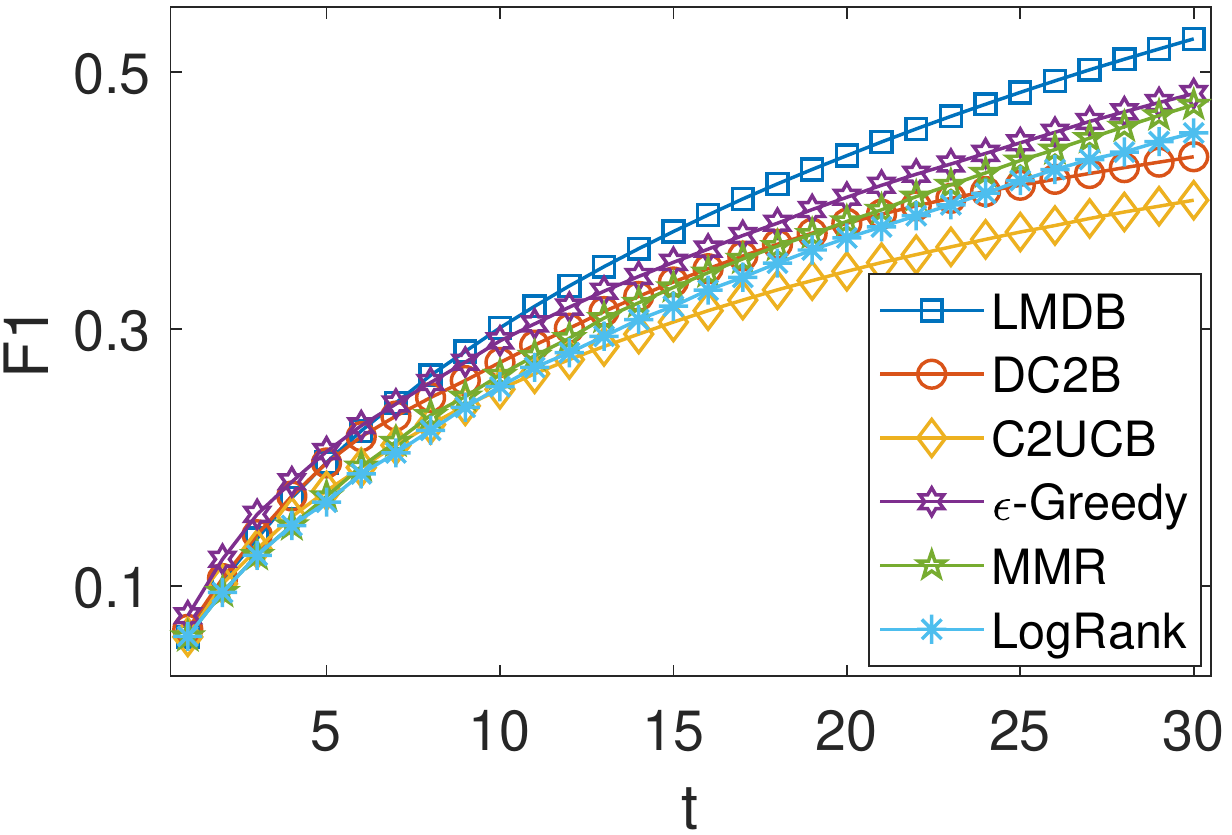}
         \caption{Movielens-1M}
         \label{fig:f1score2}
     \end{subfigure}\hfill\begin{subfigure}[b]{0.33\textwidth}
         \centering
         \includegraphics[width=\textwidth]{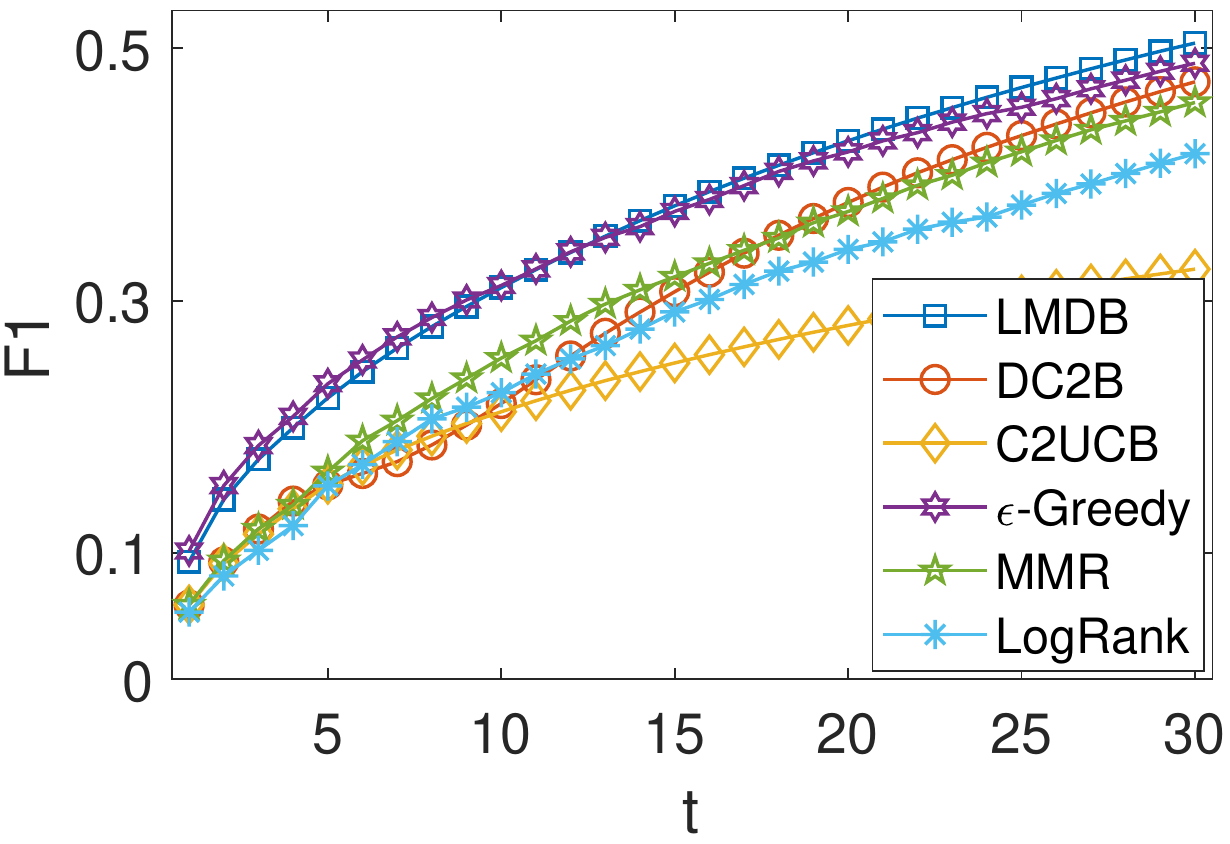}
         \caption{Anime}
         \label{fig:f1score3}
     \end{subfigure}
     \caption{The recommendation performances of different methods measured by $\mbox{F}_1$-score.}
        \label{fig:f1score}
\end{figure*}

We use Recall(t) to measure the accumulative recommendation accuracy over all the $t$ recommendation rounds, and measure the diversity of recommendation results by Diversity(t) which is defined by computing the average recommendation diversity over all the $t$ recommendation rounds~\cite{zhang2008avoiding}. We define $\mbox{Recall(t)} = \sum_{\ell=1}^t \frac{|A_\ell \bigcap \mathcal{I}|}{|\mathcal{I}|}$ and $\mbox{Div.(t)} = \frac{1}{t} \sum_{\ell=1}^t\big[\frac{2}{|A_\ell|(|A_\ell|-1)}\sum_{\{i,j\}: i,j\in A_\ell} (1-sim(\vect{z}_i, \vect{z}_j))\big]$, where $A_\ell$ is the recommended item set at the $\ell$-th round, and $\mathcal{I}$ denotes the set of items interacted with the user in testing data. We compute the accuracy and diversity for each user and report the averaged values over all users to obtain the accuracy and diversity metrics. $\mbox{F}_{\beta}$-score is used to measure the effectiveness of each method in balancing recommendation accuracy and diversity, where $\mbox{F}_{\beta}\mbox{(t)}=(1+\beta^2)\ast \mbox{Recall(t)} \ast \mbox{Diversity(t)}/[\beta^2\ast \mbox{Diversity(t)} + \mbox{Recall(t)}]$. $\beta$ is a constant indicating Recall is considered as $\beta$ times as important as Diversity. Empirically, we set $\beta$ to 1 and 2.

We compare LMDB with the following baseline methods: (1) $\mathbf{LogRank}$: In this method, the relevance score of each item $a$ is computed as $r_a=1/(1+\exp(-\bar{\vect{u}}\vect{z}_a^T))$, where $\bar{\vect{u}}$ is the mean of the user embeddings learnt from the training data. Then, this method selects $K$ items with the highest quality scores as $A_{t}$ in the $t$-th recommendation round; (2) $\mathbf{MMR}$ \cite{carbonell1998use}: For each round $t$, this method sequentially selects an available item with the largest maximal marginal score, which is defined as $\bar{r}_a=\alpha r_a-\frac{1-\alpha}{|A_t|}\sum_{j\in A_t}sim(\vect{z}_a,\vect{z}_j)$, where $r_a$ is the item quality defined in the LogRank method; (3) $\epsilon$\textbf{-Greedy}: This method randomly selects an item into the set $A_t$ with probability $\epsilon$, and selects the item with the highest relevance score into the set $A_t$ with probability $1-\epsilon$, which is defined the same as in LogRank method; (4) $\mathbf{C2UCB} $ \cite{qin2014contextual}: This is a contextual combinatorial bandit with an entropy regularizer for diversity recommendation; (5) $\mathbf{DC2B}$ \cite{liu2019bandit}: This is a full Bayesian online recommendation framework developed based on determinantal point process. We set the size of $A_t$ to 10 for all methods. The hyper-parameter settings for each method are summarized in Appendix.


\textbf{Experiment Results.} The recommendation accuracy and diversity of different methods are shown in Figure~\ref{fig1} and Figure~\ref{fig2}, respectively. As shown in Figure~\ref{fig1}, the proposed LMDB framework consistently achieves the best recommendation accuracy over all 30 recommendation rounds on all datasets. When performing more interactions between the user and the recommender agent, LMDB and C2UCB can gradually achieve more improvements over other methods. For DC2B, it is effective in the first few recommendation rounds. However, its accuracy then becomes poorer than other methods with the increase of the number of recommendation rounds. 
Moreover, from Figure~\ref{fig2}, we can note that LMDB usually achieves the second or third best recommendation diversity on all datasets. Over all recommendation rounds, LMDB consistently achieves better recommendation diversity than C2UCB, and it can usually generate more diverse recommendation results than DC2B in the first 15 rounds.

\begin{figure*}
     \centering
     \begin{subfigure}[b]{0.33\textwidth}
         \centering
         \includegraphics[width=\textwidth]{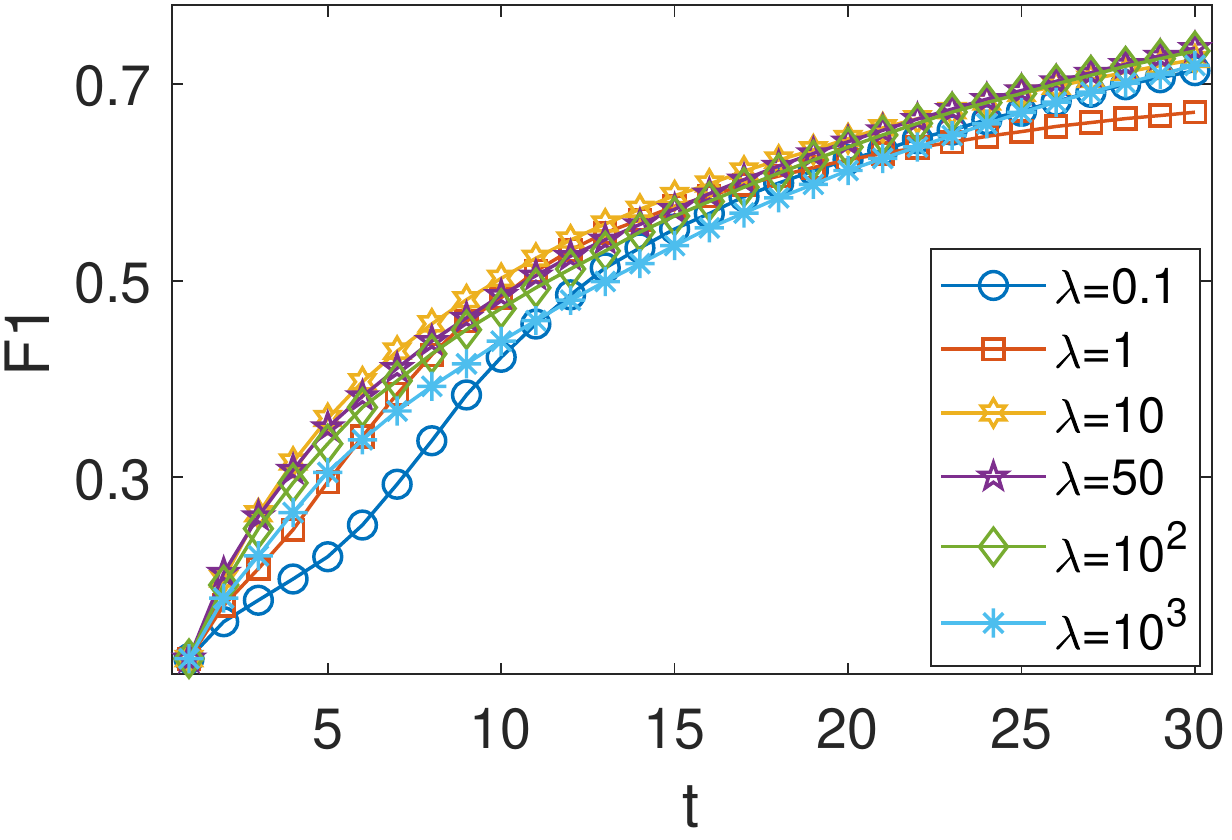}
         \caption{$\lambda$}
         \label{fig:ml100kf1score1}
     \end{subfigure}\hfill\begin{subfigure}[b]{0.33\textwidth}
         \centering
         \includegraphics[width=\textwidth]{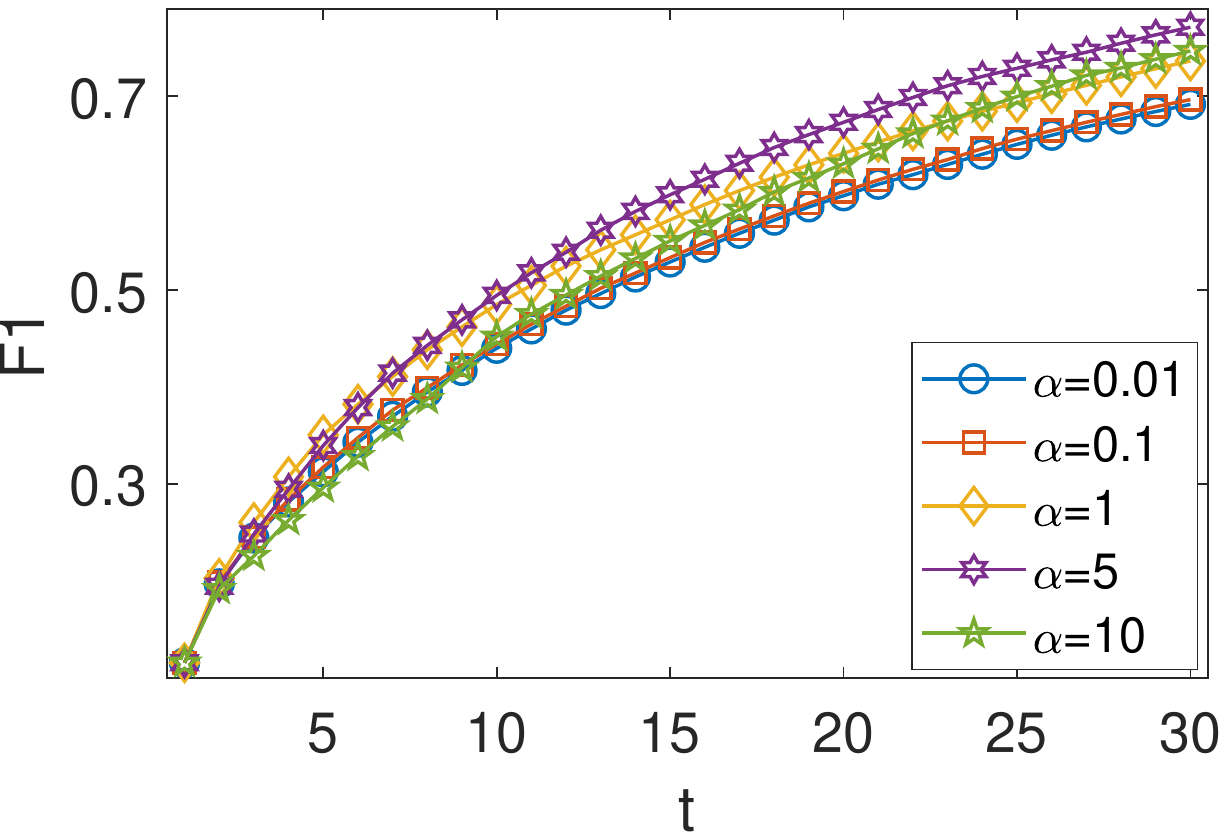}
         \caption{$\alpha$}
         \label{fig:ml100kf1score2}
     \end{subfigure}\hfill\begin{subfigure}[b]{0.33\textwidth}
         \centering
         \includegraphics[width=\textwidth]{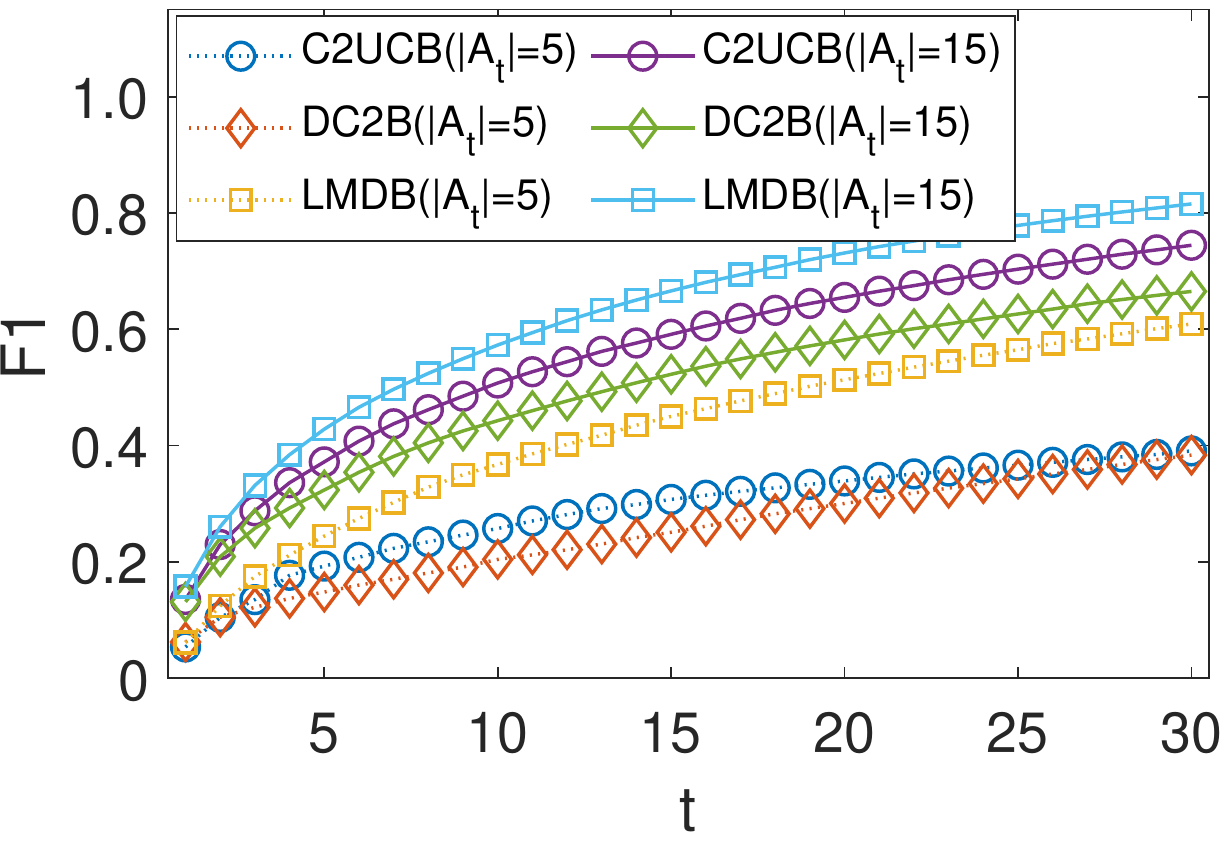}
         \caption{$|A_t|$}
         \label{fig:ml100kf1score3}
     \end{subfigure}
     \caption{The $\mbox{F}_1$-score of LMDB w.r.t. different settings of $\lambda$, $\alpha$, and $|A_t|$ on Movielens-100K.}
        \label{fig:ml100kf1score}
\end{figure*}

For better understanding of the results, $\mbox{F}_{\beta}$-score is used to study the effectiveness of each method in balancing the recommendation accuracy and diversity. Figure~\ref{fig:f1score} summarizes the $\mbox{F}_1$-score achieved by different methods. As shown in Figure~\ref{fig:f1score1} and~\ref{fig:f1score2}, LMDB usually outperforms other baselines over all recommendation rounds, in terms of $\mbox{F}_1$-score. From Figure~\ref{fig:f1score3}, we can also notice that LMDB achieves comparable $\mbox{F}_1$-score with $\epsilon$-Greedy and better $\mbox{F}_1$-score than other baseline methods. $\epsilon$-Greedy is a simple exploration method that can usually achieve high recommendation diversity. Thus, it may also have a high $\mbox{F}_1$-score. In practice, the recommendation accuracy is usually more important than recommendation diversity. Considering this factor, we also report the recommendation performances in terms of $\mbox{F}_2$-score, which treats Recall as 2 times as important as Diversity. The results shown in Figure 1 in Appendix indicate that LMDB achieves the best $\mbox{F}_2$-score on all datasets over all recommendation rounds. These results demonstrate that the proposed LMDB framework is more effective in balancing the recommendation accuracy and diversity than baseline methods, especially when we place more emphasis on recommendation accuracy than recommendation diversity. We believe this is because only LMDB learns the user's personalized preferences across the relevance properties and diversity properties of the recommended item set.

Moreover, we also evaluate the impacts of $\lambda$, $\alpha$, and the size of recommended item set $|A_t|$ on the performances of LMDB. As shown in Figure~\ref{fig:ml100kf1score1} and~\ref{fig:ml100kf1score2}, we can note that the values of $\lambda$ and $\alpha$ do not have significant impacts on the performances of LMDB, when $\lambda$ and $\alpha$ are equal to or larger than 1. In addition, we also vary $|A_t|$ in $\{5, 10, 15\}$. Considering both Figure~\ref{fig:f1score1} and~\ref{fig:ml100kf1score3}, we can note that the proposed LMDB framework consistently outperforms other combinatorial bandit learning methods C2UCB and DC2B in terms of $\mbox{F}_1$-score,with different settings of $|A_t|$. These results demonstrate that LMDB are stable and insensitive to the settings of hyper-parameters.

\begin{figure}
    \centering
    \includegraphics[width=0.9\columnwidth]{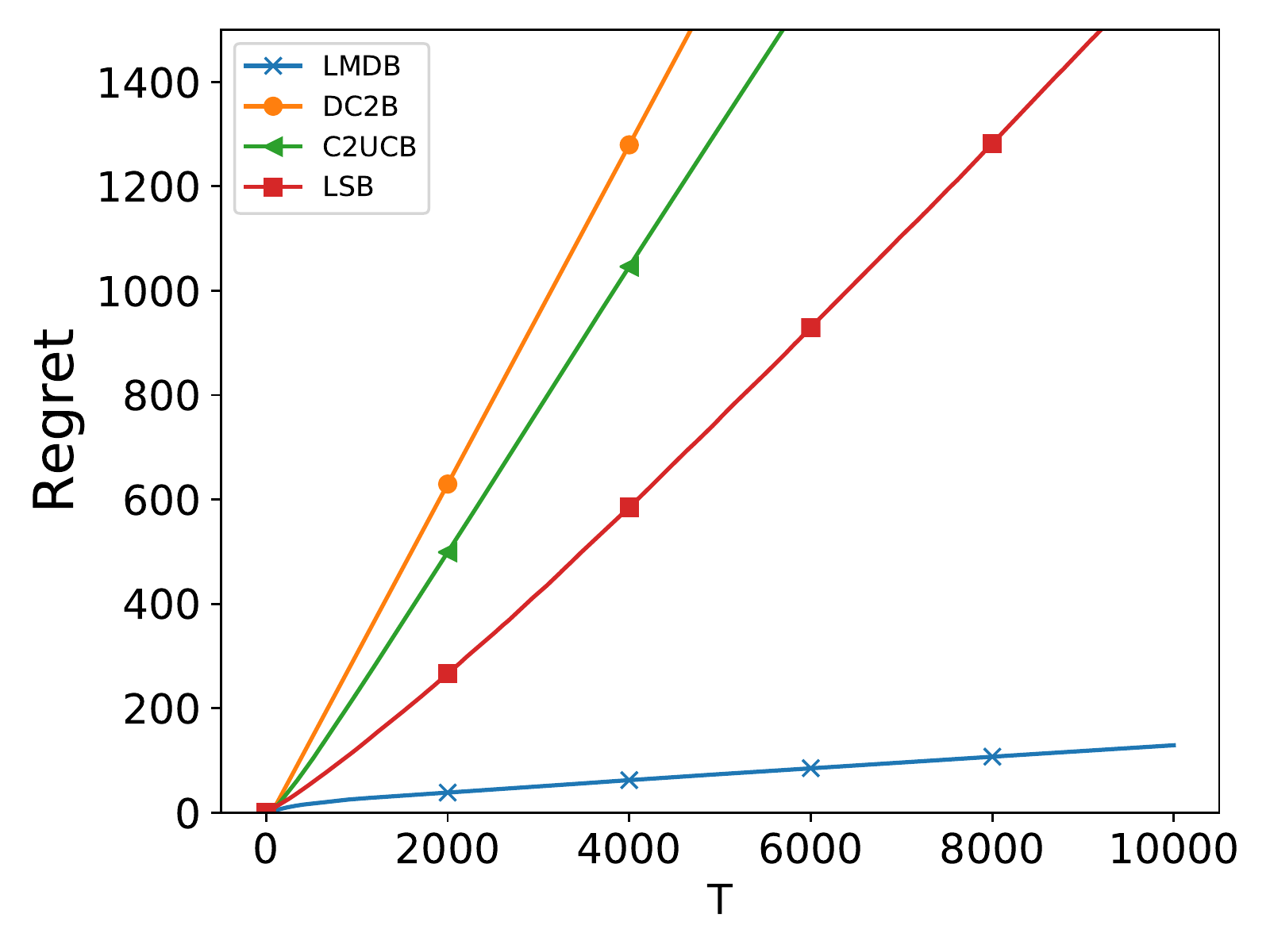}
    \caption{Regret of different bandit methods on simulated dataset.}
    \label{figregret}
\end{figure}

\subsection{Experiment on Simulated Dataset}
\label{simulated}
This experiment is to verify the regret in Theorem \ref{theorem2} and compare the result of our Algorithm \ref{alg:algorithm2} with other algorithms. Here, we only consider the bandit based methods including C2UCB, DC2B and LSB~\cite{yue2011linear}. We already describe C2UCB and DC2B in Section \ref{real-data}. LSB is a linear contextual bandit, which optimizes a class of submodular utility functions for diversified retrieval. The item's feature of LSB is extracted from the category information of the item, which is not consistent with the setting of Section \ref{real-data}. Hence, we only evaluate it for this experiment setting. We consider a setting, where there are 20 candidate items. The objective is to recommend a relevant and diverse item list to a user. The list size $K$ is empirically set to 5. The relevance features of each item are uniformly generated from $[0, 0.5]^{10}$. Here, we only use one-dimensional diversity feature. The distance metric $h(i,j)$ of the dispersion function is defined the same as in Section \ref{real-data}. The user's preferences on the relevance properties and diversity property are generated from $[0, 0.2]^{10}$ and $[0, 0.2]$, respectively. We run 20 independent simulations and compute the average regret over them.

Figure \ref{figregret} compares the performances of our algorithm with the baselines. The average regret of LMDB flattens with the number of trials $T$ and is much smaller than other baselines. Although all of the three baselines consider the diversity of the recommendation, they never learn the user's preferences on the diversity properties of the recommended item set. Hence, they recommend a diverse item set but possibly not suitable for the user.

In Section \ref{sec3}, we propose the greedy Algorithm \ref{alg:algorithm} to compute a near-optimal set for our optimization problem (Eq.~\eqref{ut}), which can reach the approximation ratio at $\frac{1}{4}$ of the optimal solution. Here, we empirically demonstrate that the approximation ratio is close to 1 in a domain of our interest. We choose 100 random users and 20 random items and vary the number of recommended items $K$ from 2 to 5. The users' preferences and items' features are generated from the uniform distribution stated above. For each user and $K$, we compute the optimal list by exhaustive search and the sub-optimal list by Algorithm \ref{alg:algorithm}. After that, we compute the approximation ratio $F(A^{greedy}|\vect{\eta})/F(A^*|\vect{\eta})$, where $F(\cdot)$ is defined in Eq. ~\eqref{ut}. We find that the average approximation ratios over all users are 0.9995, 0.9992, 0.9989, and 0.9971 when $K$ is set to 2, 3, 4, and 5 respectively. This indicates that the approximation ratio in practice may be much better than theoretical result.

\section{Conclusion and Future Work}

This paper introduces a novel bandit model named LMDB (i.e., \underline{L}inear \underline{M}odular \underline{D}ispersion \underline{B}andit) for optimizing a combination of modular functions and dispersion functions. This setting is useful for building the diversified interactive recommender systems that interactively learn the user's preferences from her timely implicit feedback. Specifically, LMDB employs modular functions and dispersion functions to describe the relevance properties and diversity properties of the recommended item list, respectively. Moreover, we propose the linear modular dispersion hybrid algorithm to solve the LMDB problem, and derive a gap-free upper bound on its scaled $n$-step regret. We conduct empirical experiments on three datasets and find that the proposed LMDB model outperforms existing bandit learning methods in balancing the recommendation accuracy and diversity. For future work, we would like to improve our analysis method to get a better approximation ratio of the optimal solution in the bandit setting.

\section{Ethical Impact}

This work has the following potential positive impacts for the companies which need recommender systems. We propose a diversified recommendation framework, which achieves good performance on the trade-off between the accuracy and diversity of recommendation results. 
It can be beneficial to show more different products to the users while keeping them in the system. At the same time, this work may have some negative consequences. The accuracy of the recommendation results would be reduced by improving the diversity of recommendation results. Thus, this will increase the risk that users may leave the system. 
Furthermore, our testing is based on an offline evaluation strategy \cite{li2011unbiased}. A direct application to online recommendation scenarios has the probability that users are not satisfied with the recommendation results. Therefore, more work is needed to test the performance of our recommendation framework on real recommendation scenarios.

\bibliography{aaai21}

\appendix
\section{Proof of Theorem 1}

We start by define some useful notations. Let $E$ be the underlying ground set, and let $\mathcal{A} \subseteq \{A\subseteq E: |A|\leq K\}$ is a family of subsets of $E$ with $K$ items. For any given subset $A\subseteq E$, $R_1(A),\dots,R_d(A)$ are modular functions corresponding to how well the recommended set of items $A$ captures the relevance properties, and $V_1(A),\dots,V_m(A)$ are dispersion functions corresponding to how well the recommended set of items $A$ captures the diversity properties. $\vect{\theta} \in \mathbb{R}^d$ is user's preferences on relevance property, and $\vect{\beta}\in\mathbb{R}^m$ is user's preference on diversity property. Then the weighted sum of relevance property is defined as
\begin{equation}
    \bar{R}(A)=\vect{\theta}^T \langle R_1(A),\dots,R_d(A)\rangle.
\end{equation}
It is obvious that $\bar{R}(A)$ is still a modular function, as the weighted sum of modular functions is still a modular function. From the definition of monotone modular function, if $\bar{R}(a)$ is non-negative for any element $a \in E$, then $\bar{R}(A)$ is a monotone modular function.
We define the weighted sum of dispersion functions as
\begin{equation}
   \bar{V}(A) =\vect{\beta}^T \langle V_1(A),\dots,V_m(A)\rangle.
\end{equation}
From the definition of dispersion function, it is obviously that $\bar{V}(A)$ is still a dispersion function if $\vect{\beta}$ is non-negative.
From Eq. (1) in the main paper, we observe the object function $F(A|\vect{\eta})$ of recommending $A$ is defined as
\begin{equation}
    F(A|\vect{\eta}) = \bar{R}(A) + \bar{V}(A).
\end{equation}
Then $F(A|\vect{\eta})$ is the combination of a monotone modular function and a dispersion function. We define another object function $ F'(A|\vect{\eta})$ as
\begin{equation}
    F'(A|\vect{\eta}) = \bar{R}(A)+\frac{1}{2}\bar{V}(A).
\end{equation}
Let $A^{greedy}$ be the greedy solution of Algorithm 1. Then from the proof of Theorem 1 of (Borodin, Lee, and Ye 2012), we have
\begin{equation}
\begin{array}{rll}
   F'(A^{greedy}|\vect{\eta}) &\geq& \displaystyle \frac{1}{2}\mathop{\arg\max}_{A\in\mathcal{A}} F'(A|\vect{\eta})\\\vspace{-2mm}\\
    &=&\displaystyle \frac{1}{2}\mathop{\arg\max}_{A\in\mathcal{A}} \bar{R}(A)+\frac{1}{2}\bar{V}(A)\\\vspace{-2mm}\\
    &\geq& \displaystyle \frac{1}{4} \mathop{\arg\max}_{A\in\mathcal{A}} \bar{R}(A)+\bar{V}(A)\\\vspace{-2mm}\\
    &=& \displaystyle \frac{1}{4} F(A^*|\vect{\eta}).
\end{array}
\end{equation}
From the definition of $F(A|\vect{\eta})$ and $F'(A|\vect{\eta})$, we have $F(A|\vect{\eta})\geq F'(A|\vect{\eta})$ for any $A\in \mathcal{A}$.
Thus, we have
\begin{equation}
    F(A^{greedy}|\vect{\eta}) \geq \frac{1}{4} F(A^*|\vect{\eta}).
\end{equation}
This concludes the proof for Theorem 1.

\begin{figure*}
     \centering
     \begin{subfigure}[b]{0.3\textwidth}
         \centering
         \includegraphics[width=\textwidth]{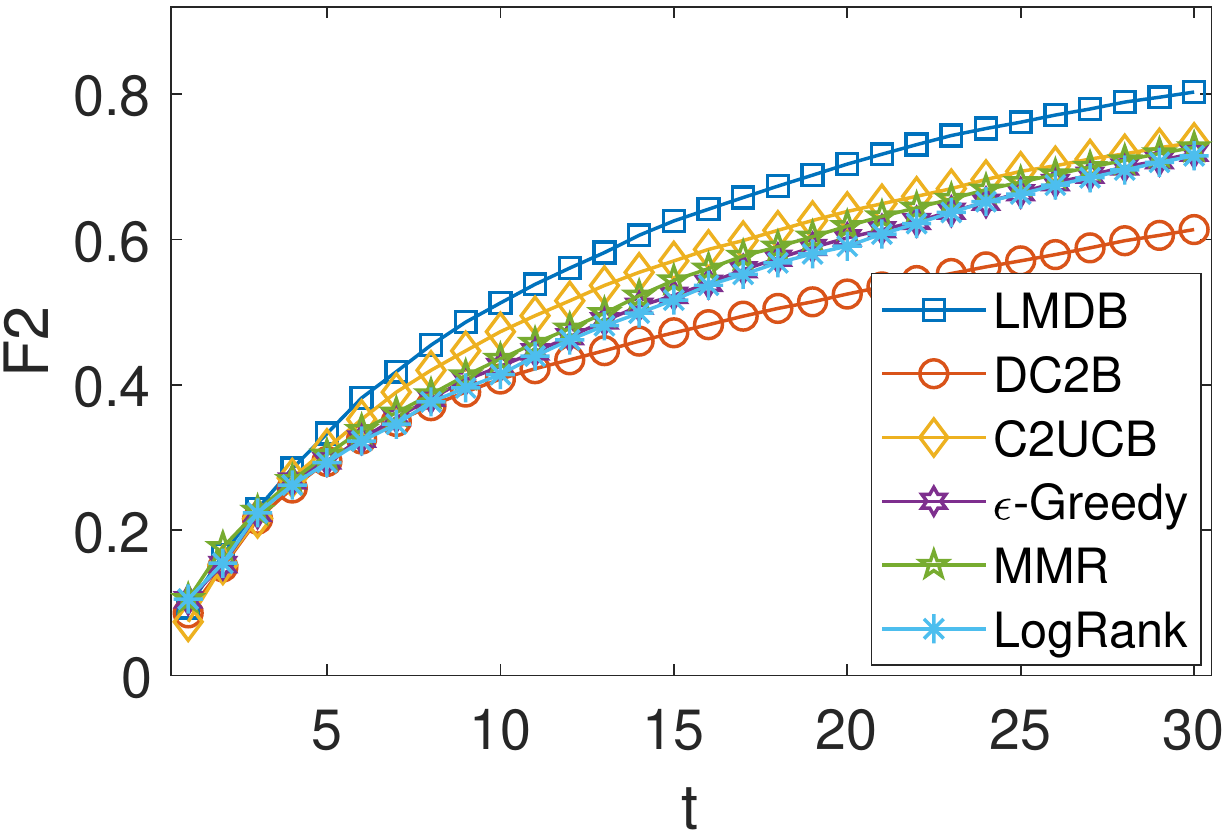}
         \caption{ML-100K}
     \end{subfigure}\hfill\begin{subfigure}[b]{0.3\textwidth}
         \centering
         \includegraphics[width=\textwidth]{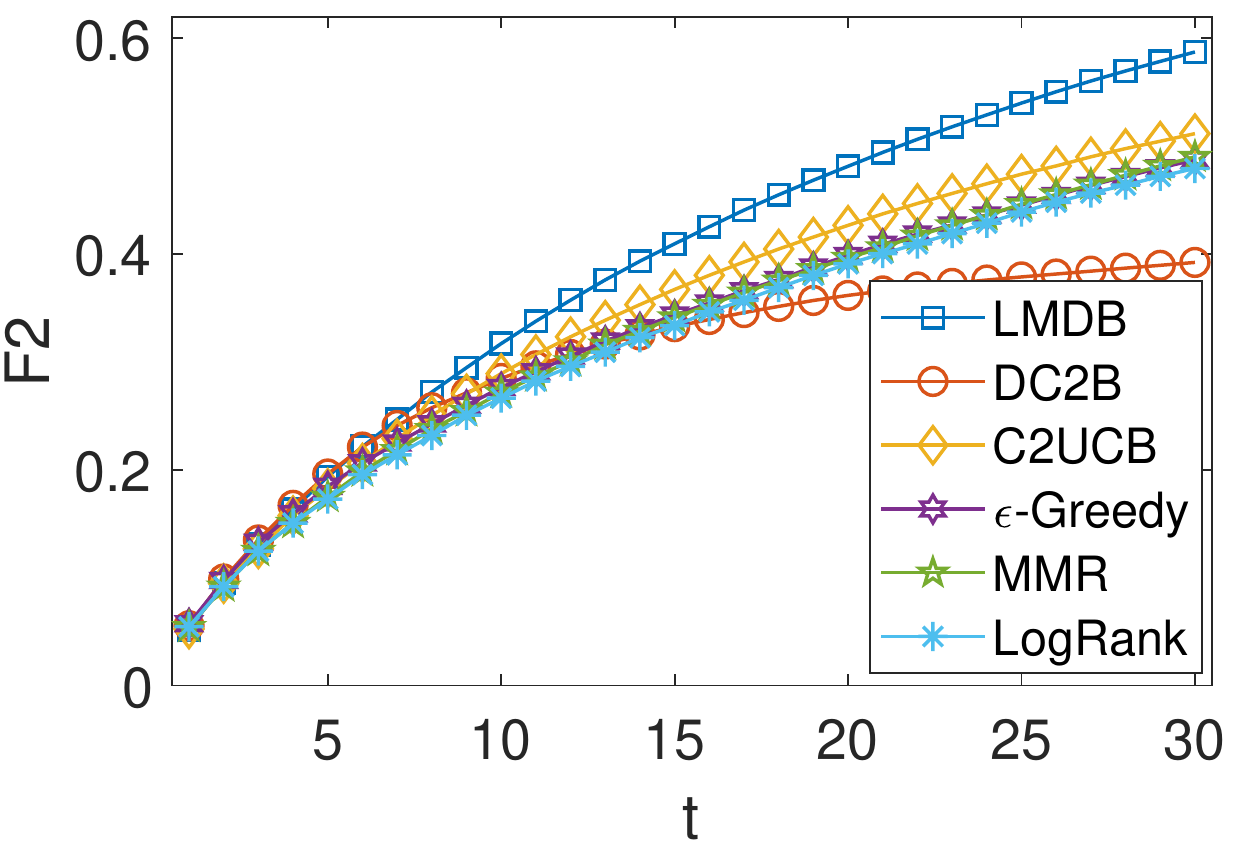}
         \caption{ML-1M}
     \end{subfigure}\hfill\begin{subfigure}[b]{0.3\textwidth}
         \centering
         \includegraphics[width=\textwidth]{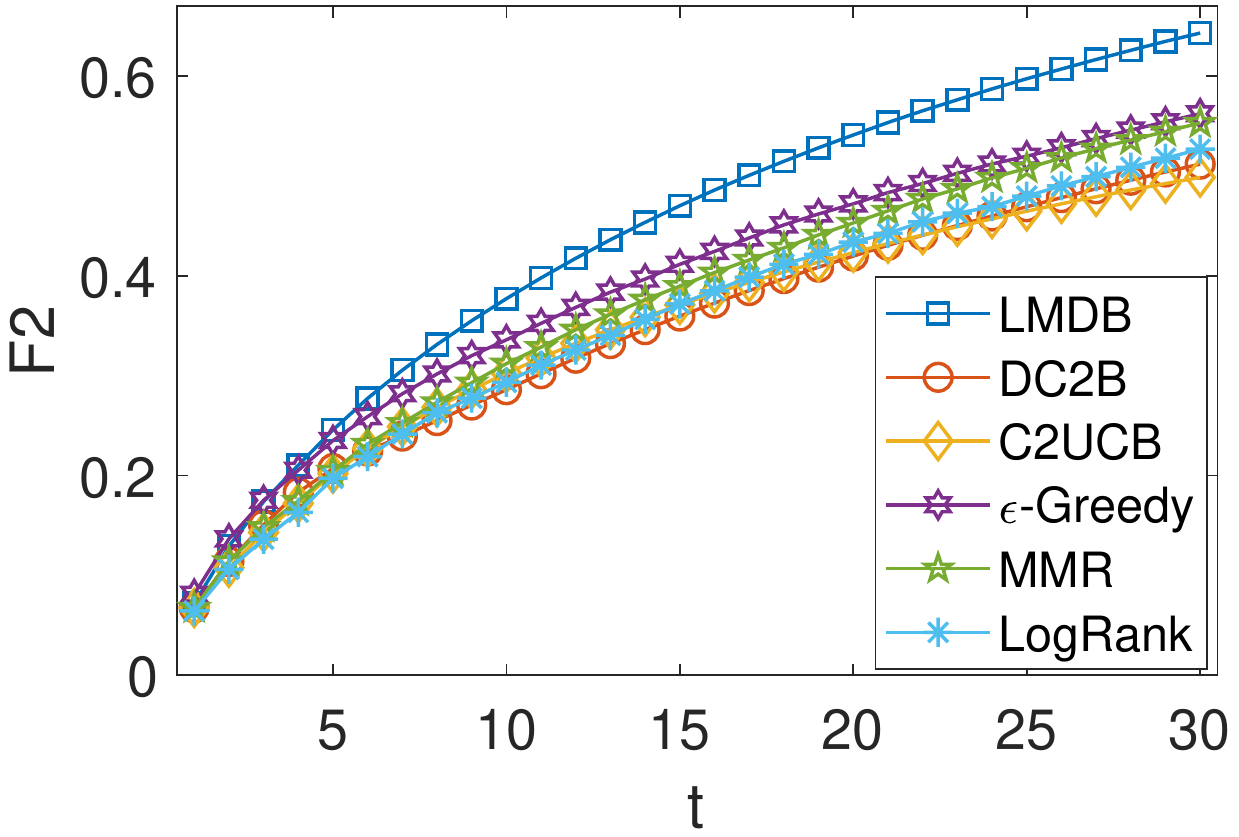}
         \caption{Anime}
     \end{subfigure}
     \caption{The recommendation performances of different methods measured by $\mbox{F}_2$-score.}
        \label{fig4}
\end{figure*}

\begin{table*}
  \caption{Statistics of the experimental datasets.}
  \label{table1}
  \centering
  \begin{tabular}{|c|c|c|c|c|c|}
    \hline
    Datasets & Users  & Items & Interactions & Categories & Density \\
    \hline
    ML-100K  & 942    & 1,447  & 55,375       & 18         & $4.06\%$ \\
    \hline
    ML-1M    & 6,038  & 3,533  & 575,281      & 18         & $2.70\%$ \\
    \hline
    Anime   & 69,400 &8,825 & 5,231,117 & 44 & $0.85\%$ \\
    \hline
  \end{tabular}
\end{table*}

\section{Proof of Theorem 2}
We start by define some useful notations. We write $\vect{\zeta}_a = [\vect{z}_a^T, \vect{x}_a^T]^T$ as the feature of item $a$. Let $\vect{\Phi}_t=\lambda\vect{I}_{d+m}+\sum_{i=1}^{t-1}\sum_{a\in A_i} \vect{\zeta}_a\vect{\zeta}_a^T$ as the collected features in t steps. Let $A^k$ be the list defined by $A^k=(a_1,a_2,\dots,a_k)$. Then, the confidence interval of Algorithm 2 on the $k$-item can be written as
\begin{equation}
  \alpha \sqrt{v_{a_k}} = \alpha\sqrt{\vect{\zeta}_{a_k}^T \vect{\Phi}_t^{-1}\vect{\zeta}_{a_k}}.
\end{equation}
Let $\Pi(E)=\bigcup_{k=1}^{L} \Pi_k(E)$ be the set of all (ordered) lists of set $E$ with cardinality $1$ to $L$, and $w: \Pi(E) \rightarrow [0,1]$ be an arbitrary weight function for lists. For any $A \in \Pi(E)$ and any $w$, we define
\begin{equation}
    f(A, w)= \sum_{k=1}^{|A|} w(A^k).
\end{equation}
For any list $A$ and time $t$, we define the weight function $\bar{w}$, its upper bound $U_t$, and its lower bound $L_t$ as
\begin{equation}
    \begin{aligned}
    \bar{w}(A) &= \vect{\eta}^{*T}\vect{\zeta}_{a_l}, \\
    U_t(A) &=  \displaystyle \mbox{Proj}_{[0,1]}\left(\hat{\vect{\eta}}^{T}_t\vect{\zeta}_{a_l}+\alpha\sqrt{\vect{\zeta}_{a_l}^T \vect{\Phi}_t^{-1}\vect{\zeta}_{a_l}}\right),\\
    L_t(A) &= \displaystyle \mbox{Proj}_{[0,1]}\left(\hat{\vect{\eta}}^{T}_t\vect{\zeta}_{a_l}-\alpha\sqrt{\vect{\zeta}_{a_l}^T \vect{\Phi}_t^{-1}\vect{\zeta}_{a_l}}\right),
    \end{aligned}
\end{equation}
where $l=|A|$ and $\mbox{Proj}_{[0,1]}(\cdot)$ projects a real number into interval $[0,1]$. It is obvious that $f(A,\bar{w})=F(A|\vect{\eta}^*)$ for all order list $A$. Let $\mathcal{H}_t$ be the history of past actions and observations by the end of time period $t$. It is obvious that the upper bound function $U_{t}$, the lower bound function $L_{t}$ and the solution $A_t$ are all deterministic conditioned on $\mathcal{H}_{t}$. For any time $t$, we define the event $\epsilon_t$ as
\begin{equation}
    \epsilon_t =\{L_t(A)\leq \bar{w}(A)\leq U_t(A), \forall A\in \Pi(E_t)\}.
\end{equation}
We also define the $\bar{\epsilon}_t$ as the complement of $\epsilon_t$. For any time $t$, we have
\begin{equation}
\begin{aligned}
      &\mathbb{E}[F(A^*_t|\vect{\eta}^*)-F(A_t|\vect{\eta}^*)/\gamma]=\mathbb{E}[f(A^*_t,\bar{w})-f(A_t,\bar{w})/\gamma]\\
      &\overset{(a)}{\leq} P(\epsilon_{t})\mathbb{E}[f(A^*_t,\bar{w})-f(A_t,\bar{w})/\gamma|\epsilon_{t}]+KP(\bar{\epsilon}_{t})\\
      &\overset{(b)}{\leq}  P(\epsilon_{t})\mathbb{E}[f(A^*_t,U_t)-f(A_t,\bar{w})/\gamma|\epsilon_{t}]+KP(\bar{\epsilon}_{t})\\
      &\overset{(c)}{\leq}  P(\epsilon_{t})\mathbb{E}[f(A_t,U_t)/\gamma-f(A_t,\bar{w})/\gamma|\epsilon_{t}]+KP(\bar{\epsilon}_{t}),
\end{aligned}
\label{eq25}
\end{equation}
where $(a)$ holds because $f(A^*,\bar{w})-f(A_t,\bar{w})/\gamma\leq K$, $(b)$ holds because under event $\epsilon_{t}$, we have $f(A, L_{t})\leq f(A,\bar{w})\leq f(A,U_{t})$ for all ordered list $A\in \Pi(E_t)$, and $(c)$ holds because $A_{t}$ is computed based on a $\gamma$-approximation algorithm, which is
\begin{equation}
f(A^*_t, U_{t}) \leq \mathop{\max}_{A\in \Pi_{K}(E_t)} f(A,U_{t})\leq f(A_t, U_{t})/\gamma.
\end{equation}
For any $\mathcal{H}_{t}$ such that $\epsilon_{t}$ holds, we have
\begin{equation}
    \begin{aligned}
         &\displaystyle \mathbb{E}[f(A_t, U_{t})- f(A_t, \bar{w})|\mathcal{H}_{t}]\\
         & = \displaystyle \mathbb{E}\left[\sum_{k=1}^{K}U_{t}(A_t^k)-\sum_{k=1}^{K}\bar{w}(A_t^k)\left|\right.\mathcal{H}_{t}\right]\\
         &\leq \displaystyle \mathbb{E}\left[\sum_{k=1}^{K}U_{t}(A_t^k)-\sum_{k=1}^{K}L_{t}(A_t^k)\left|\right.\mathcal{H}_{t}\right]\\
        &= \displaystyle 2\alpha\mathbb{E}\left[\sum_{k=1}^{K}\sqrt{\vect{\zeta}_{a_t^k}^T \vect{\Phi}_t^{-1}\vect{\zeta}_{a_t^k}}\right].\\
    \end{aligned}
    \label{eq28}
\end{equation}
Substitute Eq.~\eqref{eq28} into Eq.~\eqref{eq25}, we have
\begin{equation}
    \begin{aligned}
&\mathbb{E}[F(A^*_t|\vect{\eta}^*)-F(A_t|\vect{\eta}^*)/\gamma]\\
&\leq \frac{2\alpha}{\gamma}\mathbb{E}\left[\sum_{k=1}^{K}\sqrt{\vect{\zeta}_{a_t^k}^T \vect{\Phi}_t^{-1}\vect{\zeta}_{a_t^k}}\right]
+KP(\bar{\epsilon}_{t}).
\label{eq29}
    \end{aligned}
\end{equation}
Substitute Eq.~\eqref{eq29} into the definition of the regret (Eq.(12) of the main paper), we have
\begin{equation}
     R^\gamma(n) \leq \frac{2\alpha}{\gamma}\mathbb{E}\left[\sum_{t=1}^n\sum_{k=1}^{K}\sqrt{\vect{\zeta}_{a_t^k}^T \vect{\Phi}_t^{-1}\vect{\zeta}_{a_t^k}}\right]+nKP(\bar{\epsilon}_{t}).
\end{equation}
We need the following two lemmas to continue our proof.

\begin{lemma}
\label{lemma1}
If $\lambda \geq 0$, we have
\begin{equation}
    \sum_{t=1}^n\sum_{k=1}^{K}\sqrt{\vect{\zeta}_{a_t^k}^T \vect{\Phi}_t^{-1}\vect{\zeta}_{a_t^k}} \leq K\sqrt{\frac{n(d+m)\log \left[1+\frac{nK}{(d+m)\lambda}\right]}{\lambda\log(1+\frac{1}{\lambda})}}.
    \label{eq31}
\end{equation}
\end{lemma}

The derivation of Lemma 1 is similar with the derivation of Lemma 4 in (Wen, Kveton, and Ashkan 2015).

\begin{lemma}
\label{lemma2}
For any $t, \lambda>0, \delta \in(0,1)$ and
\begin{equation}
\begin{aligned}
    \alpha \geq  &\sqrt{(d+m)\log \left[1+\frac{nK}{(d+m)\lambda}\right]+2\log\left(\frac{1}{\delta}\right)}\\
    &+\lambda^{\frac{1}{2}}\|\vect{\eta}^*\|_2,
    \end{aligned}
\end{equation}
we have $P(\bar{\epsilon}_t)\leq \delta$.
\end{lemma}

The derivation of Lemma 2 is similar with the derivation of Lemma 4 in (Wen, Kveton, and Ashkan 2015).  Based on the above two lemmas, if we choose
\begin{equation}
\begin{aligned}
    \alpha \geq  &\sqrt{(d+m)\log \left[1+\frac{nK}{(d+m)\lambda}\right]+2\log\left(\frac{1}{\delta}\right)}\\
    &+\lambda^{\frac{1}{2}}\|\vect{\eta}^*\|_2,
    \end{aligned}
\end{equation}
we have $P(\bar{\epsilon}_t)\leq \delta$ for all $t$ and hence
\begin{equation}
     R^\gamma(n) \leq \frac{2\alpha K}{\gamma}\sqrt{\frac{n(d+m)\log \left[1+\frac{nK}{(d+m)\lambda}\right]}{\lambda\log(1+\frac{1}{\lambda})}}+nK\delta.
\end{equation}
This concludes the proof for Theorem 2.

\section{Experiments}

On Movielens-100K and Movielens-1M, all users in testing data are used for model evaluation, and the set of all items in training data are used as the initial candidate item set. For Anime dataset, we construct the initial candidate item set by choosing 1,000 most popular items in the training data, and we only test the performances of different models on the 10,000 most active users in the testing data.

The best parameter settings for each method are as follows. In MMR, the $\alpha$ is set to 0.9. In $\epsilon$-Greedy, the $\epsilon$ is set to 0.05. In $\mbox{C2}\mbox{UCB}$, we set $\lambda_0=100,\lambda=0.1,$ and $\sigma=1$. In $\mbox{DC2}\mbox{B}$, we set $\alpha=3$ and $\lambda=1$. In LMDB, we set $\lambda=50$ and $\alpha=1$. In LSB, we set $\lambda=100$ and $\alpha=1$. These parameter settings are used for experiments on real-world datasets and simluated datasets.

In practice, the accuracy is usually more important than diversity. Therefore, the F2 score is a reasonable metric to explain the balance between accuracy and diversity. In Figure \ref{fig4}, we plot the F2 score for different methods. From Figure \ref{fig4}, we observe LMDB outperforms other baseline methods, which indicates that LMDB can perform better on the trade-off between the accuracy and diversity of recommendation results.

\end{document}